\documentclass[twocolumn]{aastex631}
\usepackage{orcidlink}
\usepackage{amsmath}
\usepackage{comment}

\newcommand\St{{\rm St}}
\newcommand{\red}[1]{\textcolor{black}{#1}}
\usepackage{booktabs} 
\usepackage{float}
\usepackage{graphicx}
\usepackage{hyperref}
\usepackage{bm}
\usepackage{xcolor}

\begin{document}

\title{\Large ``Ashfall" Induced by Molecular Outflow in Protostar Evolution $\rm I\hspace{-1.2pt}I$: \\ \red{Analytical Study on the Maximum Size of Dust Grains Lifted by Outflows}  }

\author[0009-0000-9609-5953]{Hayato Uchimura}
\affiliation{Graduate School of Science and Engineering, Kagoshima
  University, 1-21-35 Korimoto, Kagoshima, Kagoshima 890-0065, Japan}

\author[0000-0003-1292-940X]{Takahiro Kudoh}
\affiliation{Faculty of Education, Nagasaki University, 1-14 Bunkyo-machi, Nagasaki 852-8521, Japan}

\author[0000-0001-6738-676X]{Yusuke Tsukamoto}
\affiliation{Graduate School of Science and Engineering, Kagoshima
  University, 1-21-35 Korimoto, Kagoshima, Kagoshima 890-0065, Japan}

\begin{abstract}
In this paper, we study the dust dynamics in the molecular outflow using an analytical magnetohydrodynamical outflow model. Specifically, we investigate the maximum size of dust grains $a_{\rm d,max}$ that can be lifted by the outflow and whether they can escape into interstellar space. We also investigate the dependence of the maximum size of the dust grains on various outflow parameters, such as the mass ejection rate of the outflow $\dot{M}$, the disk size $r_{\rm disk}$, the mass of the central protostar $M_{*}$ and the internal dust density $\rho_{\rm mat}$. 
We find the empirical formula for the maximum dust size as a function of the outflow parameters. $a_{\rm d,max}$ depends on the parameters as $a_{\rm d,max}$ $\propto$ $\dot{M}^{1.0}$ $r_{\rm disk}^{-0.44}$ $M_{*}^{-0.82}$ $\rho_{\rm mat}^{-1.0}$. We also find that the dust grains with size of 100 $\rm \mu$m to 1 mm can be lifted by the outflow and can distribute in the well outside of the disk when the mass ejection rate of the outflow has the value of $10^{-7}\,M_{\odot}\rm \,yr^{-1}$ to $10^{-6}\,M_{\odot}\rm \,yr^{-1}$.
This result is consistent with recent observations that show a correlation between the mass ejection rate of the outflow and the spectral index $\beta$ of the dust opacity at the envelope scale. 
\end{abstract}

\keywords{magneto-hydrodynamical outflow, dust lifting, millimeter-sized dust grains, envelope, Class 0/I young stellar objects}

\section{Introduction} \label{sec:intro}
The size of the dust grains in a protoplanetary disk can be estimated by examining the spectral index $\beta$ of the dust opacity in (sub)millimeter wavelengths \citep{2001ApJ...553..321D,2006ApJ...636.1114D,2010A&A...512A..15R}. Recent multi-wavelength observations of dust thermal emission have reported a decrease in the spectral index $\beta$ of dust in the early stages of protostellar evolution. This is evidence that dust growth is proceeding at an early stage in the evolution of the disk \citep{2015ApJ...813...41P,2016A&A...588A..53T,2019ApJ...883...71C}. 
The dust growth timescale is several thousand years within the disk, and from a theoretical point of view, it is also plausible that dust growth proceeds in the disk \citep{2021ApJ...920L..35T}.
 
On the other hand, recent observations have suggested the existence of grown dust in the envelope, which is difficult to understand based on the conventional view of dust growth. Several observations have shown that the spectral index $\beta$ of the dust opacity in the envelopes of young protostars in the Class 0/I phase is noticeably lower than the typical value found in the interstellar medium (ISM), as is the case in the disks. The typical value for the ISM is $\beta \sim 1.7$ \citep{1999ApJ...524..867F,2001ApJ...548..296W,2014A&A...566A..55P,2014A&A...571A..11P}. This strongly suggests that the (sub)millimeter-sized dust grains exist in this environment \citep{2007ApJ...659..479J,2009ApJ...696..841K,2015ApJ...808..102K,2012ApJ...756..168C,2014A&A...567A..32M,2019A&A...632A...5G,2019MNRAS.488.4897V}. 
This is very challenging to explain from a theoretical perspective. This is because the timescale for the dust grains to grow to $\sim 1$ mm in the envelope is nearly $100$ times longer than the free fall time of the envelope \citep[see B2 of][]{2021ApJ...920L..35T}, so it is extremely difficult for the dust grains to grow to a size of around $1$ mm in the envelope before they accrete to the disk. Dust growth simulations in envelopes also show that it is difficult to form (sub)millimeter-sized dust grains within envelopes \citep{2009A&A...502..845O,2016PASJ...68...67W,2020A&A...643A..17G,2022ApJ...940..188S,2023MNRAS.518.3326L}.
This has given rise to a tension between observational results and theoretical research.

One mechanism that could resolve the tension between observation and theory is a model in which the dust grains grown within the disk are supplied to the envelope by magnetically driven outflows.
Such a model was discovered due to the recent development of simulation techniques, making it possible to perform multi-dimensional simulations of dust-gas two-fluid \citep[][]{2019A&A...631A...1V,2019A&A...627A.154V}, and in particular, magneto-hydrodynamics simulations \citep{2020A&A...641A.112L,2021ApJ...913..148T,2021ApJ...920L..35T}. \par

In particular, \citet{2021ApJ...920L..35T} investigated the three-dimensional motion of the grown dust around the protostar using three-dimensional magneto-hydrodynamics simulations that considered the dust growth in the disk.
They found that the dust grains that had grown to several mm to cm inside the disk were lifted by the magnetically driven outflow from the upper layer of the disk and were ejected from the disk. They also pointed out that the grown dust grains decouple from the gas as the outflow gas becomes dilute, and are supplied to the envelope. The grown dust grains eventually reaccrete to the outer region of the disk. This chain of processes is named as the “Ashfall" phenomenon and is a plausible mechanism for explaining the existence of the $\gtrsim 1$ mm sized dust grains in the envelopes.

This model is interesting because it naturally explains the recent observations by \citet{2024ApJ...961...90C}. They show that there is a negative correlation between the mass ejection rate of the outflow and the spectral index $\beta$ in the envelopes, meaning that more powerful outflows are associated with lower values of $\beta$ in the envelopes.

While three-dimensional simulations are realistic,  they can only investigate the motion of the dust grains in a limited parameter space due to their huge computational cost. For this reason, it is difficult under various outflow parameters to quantitatively investigate, for example, the maximum size of dust grains that can be lifted by the outflow.

In order to overcome this problem, the analytical calculation that can comprehensively investigate a wide parameter space with low computational cost is an important tool.
Based on this idea, in this paper, \red{we investigate the maximum size of dust grains that can be \red{lifted up} by the outflows ($a_{\rm d,max}$) and its parameter dependence, and whether they can escape into interstellar space.}

This paper is organized as follows. In section 2, we explain the calculation methods and the details of the outflow model. We also describe how to analyze the dust dynamics in the outflow. In section 3, we describe the results. 
Finally, in section 4, we summarize and discuss our results.

\section{Methods and Models} \label{sec:style}    
\subsection{Basic Equations for the Outflow Model}
For the analytical outflow model, we solve the steady axisymmetric MHD equations along the magnetic field lines of \citep{1997ApJ...474..362K}. 

The basic equations along the magnetic field lines in the outflow are given as,
\begin{equation}
    P=K \rho^{\gamma},
\label{eos}
\end{equation}
\begin{equation}
    \rho v_{ p}=\lambda B_{ p},
\label{2}
\end{equation}
\begin{equation}
    (v_{\phi}-\Omega r)B_{p}=v_{p} B_{\phi},
\label{3}
\end{equation}
\begin{equation}
    r \bigg( v_{\phi} -\frac{B_{\phi}}{4\pi \lambda} \bigg)=L,
\label{4}
\end{equation}
\begin{equation}
    \frac{1}{2} v^{2}_{p} + \frac{1}{2} v^{2}_{\phi} + \frac{\gamma}{\gamma-1} \frac{P}{\rho} + \Psi_{\rm g} -\frac{r \Omega B_{\phi}}{4 \pi \lambda}=E.
\label{5}
\end{equation}
$P$, $v$, $B$, and $\rho$ are the gas pressure, gas velocity, magnetic field, and gas density, respectively. Subscripts $p$ and $\phi$ denote the poloidal and toroidal component, respectively. $K$, $\lambda$, $L$, $E$, and $\Omega$ are the constants along the magnetic field lines. $\gamma=1.05$ denotes the polytropic exponent. $\Psi_{\rm g}$ is the gravitational potential.  

Equation (\ref{eos}) denotes the polytropic equation of state. Equation (\ref{2}) denotes that the poloidal component of the velocity and the magnetic field are parallel in the poloidal plane under the assumption of the steady and axisymmetric flow. Equation (\ref{3}) denotes the frozen-in condition. Equation (\ref{4}) denotes the conservation of the total angular momentum. Equation (\ref{5}) denotes the conservation of energy along the magnetic field lines. In addition, following conservation equations are assumed,
\begin{equation}
    B_{p} \Sigma = \Phi=\rm const,
\label{6}
\end{equation}
\begin{equation}
    \rho v_{p}\Sigma=\dot{M}=\rm const.
\label{7}
\end{equation}
Here, $\Phi$, $\dot{M}$ are the magnetic flux and the mass flux respectively. 
$\Sigma$ is a cross section of the flux tube as a function of $r$ and is assumed to be $\Sigma=\pi r^{2}$ in this paper where $r$ is the cylindrical radius. 
Equation (\ref{6}) gives the absolute value of the poloidal magnetic field strength $B_{p}$. Equation (\ref{7}) relates the outflow rate to the density. 
We also define equation (\ref{8}) which is obtained from the condition that toroidal component of the magnetic field and velocity dot not diverge at the Alfv$\acute{\rm e}$n point,
\begin{equation}
    L=\Omega r^{2}_{\rm A}.
\label{8}
\end{equation}
$r_{\rm A}$ is the Alfv$\acute{\rm e}$n radius which is defined as where the poloidal velocity equals $B_{p}/(4\pi\rho)^{1/2}$.
The geometry of the magnetic field lines are assumed to be \citep{1994A&A...287...80C},
\begin{equation}
    [(r/r_{\rm N})^{2})+(1+\zeta)^{2}]^{\frac{1}{2}}-(1+\zeta)=\rm const,
\label{geo}
\end{equation}
\begin{equation}
    \zeta=(z/r_{\rm N})\tanh (z/H_{\mathrm{B}}).
\end{equation}
With this equation, we determine $r$ from given $z$ where $z$ is the cylindrical height of the magnetic field line. $r_{\rm N}$ is a parameter that determines the angle between the disk surface and the magnetic field line. 
We set $r_{\mathrm{N}}/r_{0}=2.5$ for our fiducial model. With this value, the inclination angle between the magnetic field line and the disk surface $\Theta$ is $\Theta=53^\circ$. 
$\zeta$ is a dimensionless quantity that includes a factor of $\tanh (z/H_{\mathrm{B}})$ so that the poloidal magnetic field line threads the midplane perpendicularly. 

$H_{\mathrm{B}}$ is 
the parameter that determines the approximate height at which the poloidal magnetic field lines begin to bend. Since, for the gas to be accelerated by magneto-centrifugal forces, the angle of the magnetic field must be $\lesssim 60^\circ $ \citep{1982MNRAS.199..883B}, $H_{\mathrm{B}}$  can be regarded as a parameter that controls the height of the launching point of the outflow.
Note that, since gas pressure is taken into account in our model, the gas is lifted mainly by the gas pressure until it reaches the saddle point of the effective potential.
Note also that, since the required magnetic field inclination decreases as $z$ increases to pass the saddle point \citep[see, Fig. 1 of][]{1982MNRAS.199..883B}, an outflow solution can be obtained even for the angle slightly larger than $\Theta=60^\circ$ (section \ref{sec_inc}).

We set $H_{\rm B}=0.1r$ except for section \ref{z0_dependence}. 
In our model, the wind base is set to be $z=0$ and we control the height around which the gas begins to be accelerated by changing $H_{\rm B}$. In contrast, \citet{2016ApJ...818..152B} assumes the straight inclined magnetic field and wind base is set to be $z=4H$ where $H$ is the scale height.

There are six unknown quantities $\rho$, $v_{p}$, $B_{p}$, $v_{\phi}$, $B_{\phi}$ and $P$. By specifying these six parameters at the wind base $\rho_{0}$, $v_{p0}$, $B_{p0}$, $v_{\phi0}$, $B_{\phi0}$ and $P_{0}$ (where the subscript 0 denotes the value at the wind base), the constants along the magnetic field line $K$, $\dot{M}$, $\Omega$, $\Psi$, $L$, $E$ are determined.  By solving equation (\ref{eos}), (\ref{3})-(\ref{5}), (\ref{6}) and (\ref{7}) with the geometry of equation (\ref{geo}), we can obtain a wind solution. 
However, there are additional constraints that the solutions should pass through two critical points (fast, slow). As a result, there are two additional equations, and we can only specify four quantities at the wind base out of the six quantities. 
Below, we explain these relations using non-dimensional form.

\begin{figure}[t]
         \centering
         
         \hspace{-0.2cm}
    \includegraphics[scale=0.58]{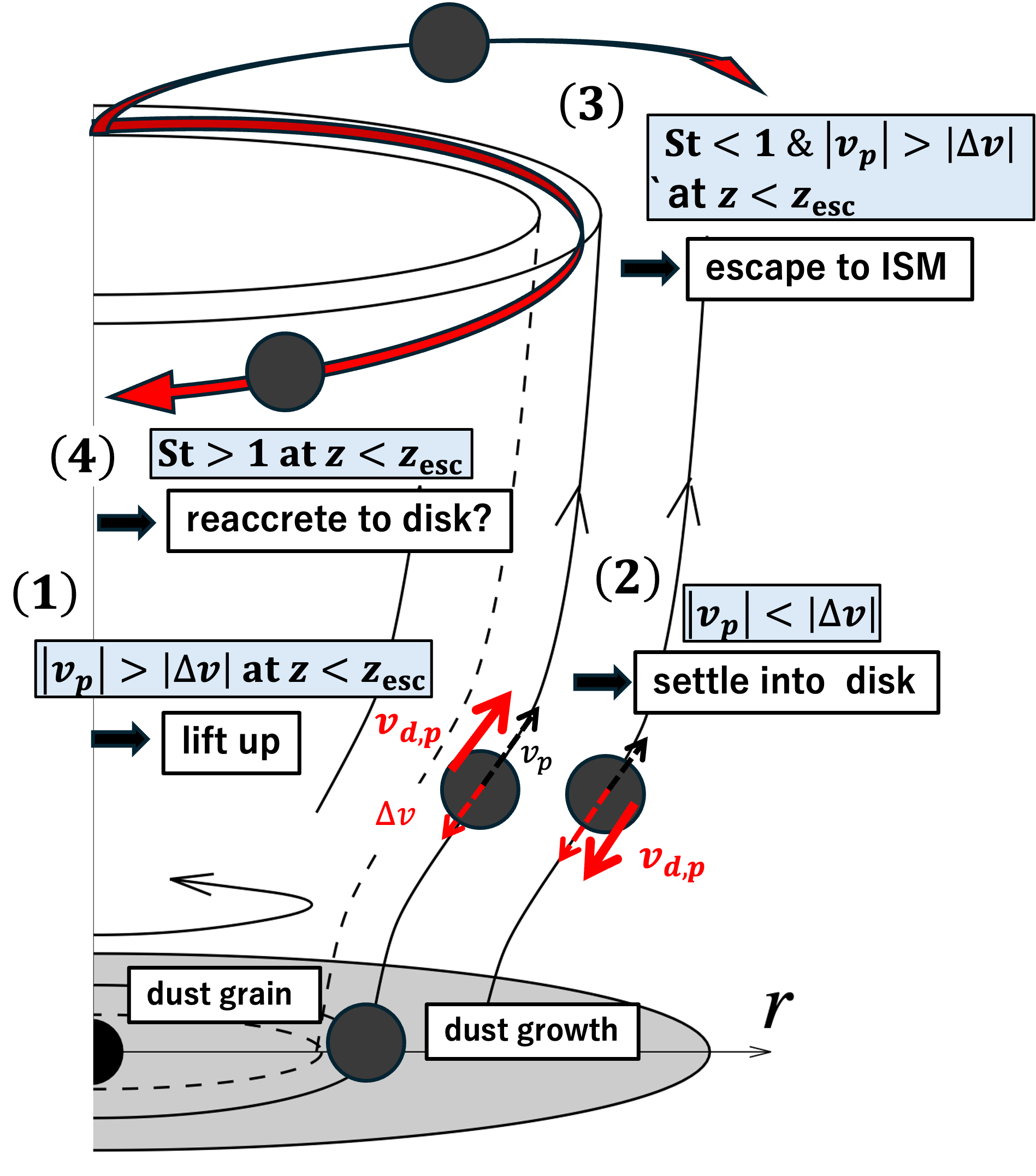}
    \caption{
    Schematic figure of this study. In this study, the dust grains are assumed to grow in the disk. \red{The dust grains \red{are assumed to be} lifted up by the outflow if the poloidal dust velocity $v_{d,p}$ satisfies $v_{d,p}>0$, i.e., $|v_{p}|>|\Delta v|$  (1). If $v_{d,p}$ satisfies $v_{d,p}<0$, the dust grains settle into the disk (2). We assume that the dust grains escape into the interstellar space if the conditions of ($\rm St<1$) and $v_{d,p}>0$ at $z<z_{\rm esc}$ are satisfied (3). On the other hand, if the condition $\rm St>1$ at $z<z_{\rm esc}$ is satisfied, which means that the terminal velocity approximation is invalid before the dust grains reach escape velocity, they might reaccrete to the disk. }}

    \label{model1}
\end{figure}

\subsection{Equation of Non-dimensional Form}
We define the energy $\tilde{E}$ of the frame rotating with the angular velocity $\Omega$ instead of equation (\ref{5}), 
\begin{equation}
    \begin{split}
        \tilde{E} &=E-\Omega L \\ 
    &= \frac{1}{2} v^{2}_{p} + \frac{1}{2} (v_{\phi} - \Omega r)^{2} +\frac{\gamma}{\gamma-1} \frac{P}{\rho} + \Psi_{\rm g} -\frac{1}{2} \Omega^{2} r^{2}.
    \end{split}
\label{11}
\end{equation}
By substituting equation (\ref{eos}), (\ref{3})-(\ref{4}), (\ref{6}) and (\ref{7}) into equation (\ref{11}), we can obtain the equation which only involves the density. We rewrite this equation as a non-dimensional form as in \cite{1985A&A...152..121S,1987PASJ...39..821S},
\begin{equation}
\begin{split}
    h(x,y; \beta,\theta,\omega) \equiv \frac{\beta}{2y^{2}} \bigg(\frac{\Sigma_{\rm A}}{\Sigma}\bigg)^{2} + \frac{\theta}{\gamma -1} y^{\gamma-1} + \hat{\Psi}_{\rm g} \\
    +\frac{\omega}{2}\bigg[\frac{(x-1/x)^{2}}{(y-1)^{2}}-x^{2}]=e,
\end{split}   
\label{12}
\end{equation}
where
\begin{equation}
    x=\frac{r}{r_{\rm A}}, \quad   y=\frac{\rho}{\rho_{\rm A}}, \quad \hat{\Psi}_{\rm g}=\frac{\Psi_{\rm g}}{GM_{*}/r_{\rm A}},
\end{equation}
\begin{equation}
    e=\frac{\tilde{E}}{GM_{*}/r_{\rm A}}, \quad \beta=\bigg(\frac{V^{2}_{\mathrm{A},p}}{GM_{*}/r_{\rm A}} \bigg),
\end{equation}
\begin{equation}
    \theta=\bigg(\frac{a^{2}_{\rm A}}{GM_{*}/r_{\rm A}} \bigg), \quad  \omega=\bigg(\frac{\Omega^{2}r^{2}_{\rm A}}{GM_{*}/r_{\rm A}} \bigg),
\end{equation}
\begin{equation}
    a=(\gamma P/\rho)^{1/2},  \quad  V_{\mathrm{A}, p}=(B^{2}_{p}/4\pi \rho)^{1/2}.
\end{equation}
Subscript A indicates the value at the Alfv$\acute{\rm e}$n point.
$\theta$, $\beta$, $\omega$, and $e$ show the non-dimensional thermal energy, magnetic energy, rotational energy, and total energy which are normalized by the gravitational potential at the Alfv$\acute{\rm e}$n point. 
$a$ is the sound speed. To satisfy the condition that the wind solution pass through two critical points (fast, slow), it is necessary to fulfill the following equation (\ref{17}),

\begin{equation}
    \frac{\partial h}{\partial x}=\frac{\partial h}{\partial y}=0, \quad h=e.
\label{17}
\end{equation}

With dimensionless variables, there are eight parameters to be obtained $x_{\rm s}$, $x_{\rm f}$, $y_{\rm s}$, $y_{\rm f}$, $\beta$, $\theta$, $\omega$, $e$. By specifying two variables, for example, $\theta$ and $\omega$, the remaining six parameters can be determined as a function of $\theta$ and $\omega$ and then we can obtain the wind solution. Also, we can obtain the  remaining quantities at the wind base after obtaining eight variables by using following relations at the wind base,
\begin{equation}
\begin{split}
    \frac{\beta^{\gamma-1}\theta}{\omega^{4/3-\gamma}} \bigg(\frac{\Sigma_{\rm A}}{r_{\rm A}^{2}}\bigg)^{2(\gamma-1)} 
    =\frac{\gamma P_{0}/\rho_{0}}{(GM_{*}/r_{0})^{2\gamma-4/3}} \\ \times 
    \frac{(B_{p0}^{2}/4\pi\rho_{0})^{\gamma-1}}{(\Omega r_{0})^{2(4/3-\gamma)}}
    \bigg(\frac{\Sigma_{0}}{r_{0}^{2}}\bigg)^{2(\gamma-1)},
\end{split}
\label{18}
\end{equation}
\begin{equation}
    \frac{e}{\omega^{1/3}}=
    \frac{
        \frac{v_{p0}^{2}}{2} + \frac{(v_{\phi 0} - \Omega r)^{2}}{2} + \frac{\gamma}{\gamma-1} \frac{P_{0}}{\rho_{0}} - \frac{GM_{*}}{r_{0}} - \frac{\Omega^{2} r_{0}^{2}}{2}
    }{
        (GM_{*} \Omega)^{2/3}
    }.
\label{19}
\end{equation}
First two terms in the numerator of the equation (\ref{19}) can often be ignored because the poloidal velocity $v_{p0}$ is small and the toroidal velocity $v_{\phi 0}$ is nearly equal to the Keplerian velocity at the wind base in many cases.
If we assume a Keplerian disk and neglect the first two terms in the numerator of the equation (\ref{19}), these two equations (\ref{18}) and (\ref{19}) can be expressed as,
\begin{equation}
    \frac{\beta^{\gamma-1}\theta}{\omega^{4/3-\gamma}} \bigg(\frac{\Sigma_{\rm A}}{r_{\rm A}^{2}}\bigg)^{2(\gamma-1)} 
    =\gamma E_{\rm th} E_{\rm mg}^{\gamma-1} \bigg(\frac{\Sigma_{\rm 0}}{r_{\rm 0}^{2}}\bigg)^{2(\gamma-1)},
\label{20}
\end{equation}

\begin{equation}
    \frac{e}{\omega^{1/3}}=\frac{\gamma}{\gamma-1}E_{\rm th}-\frac{3}{2}.
\label{21}
\end{equation}
$E_{\rm th}$ and $E_{\rm mg}$ denote thermal energy and poloidal magnetic field energy, respectively and can be expressed as,
\begin{equation} 
\small
    E_{\rm th}=\frac{(a_{0}/V_{\rm K0})^{2}}{\gamma} = \red{7.8} \times 10^{-3} \bigg(\frac{  T_{0}}{\red{30} \rm  K} \bigg)  \bigg(\frac{ M_{*}}{\red{0.3} M_{\odot}}\bigg)^{-1} \bigg(\frac{r_{0}}{\rm \red{20}AU} \bigg),
\label{Eth}
\end{equation}
\begin{equation}
\small
\begin{split}
    E_{\rm mg}=\bigg(\frac{V_{\mathrm{A},p0}}{V_{\rm K0}}\bigg)^{2} = \red{2.5 \times 10^{-7}} \bigg(\frac{ B_{ p0}}{\red{1.3\times10^{-3}} \rm G} \bigg)^{2} \bigg(\frac{ M_{*}}{\red{0.3}  M_{\odot}} \bigg)^{-1} \\
    \times \bigg(\frac{\rho_{0}}{3.0\times10^{-12}\rm g\,cm^{-3}} \bigg)^{-1} \bigg(\frac{r_{0}}{\red{20}\rm AU} \bigg).
    \end{split}
    \label{Emg}
\end{equation}

These values are normalized by the gravitational potential at the wind base of the outflow.
The values of the thermal energy $E_{\rm th}$ and the poloidal magnetic field energy $E_{\rm mg}$ are determined by  specifying the parameters such as the temperature $T_{\rm 0}$, the radial distance from the central protostar $r_{0}$, the density $\rho_{0}$, the poloidal magnetic field strength $B_{ p0}$ at the wind base, and the mass of the central protostar $M_{*}$.
\subsection{Dust Dynamics in the Outflow}

\subsubsection{\red{Definition of the Stopping Time}}
Dust stopping time $t_{\rm stop}$ is given as,
\begin{equation}
    t_{\rm stop}=\frac{\rho_{\rm g}\rho_{\rm d}}{\rho_{\rm g}+\rho_{\rm d}}\frac{1}{K},
\label{stop}
\end{equation}
where,
\begin{equation}
\begin{split}
     K=\frac{4}{3} \pi \rho_{\rm g}  a_{\rm d}^{2} v_{\rm therm} n_{\rm d} \sqrt{1+\frac{9\pi}{128} \bigg(\frac{\Delta V}{a} \bigg)^{2}}.
\end{split}
\label{K}
 \end{equation}
 $\rho_{\rm g}$, $\rho_{\rm d}$, $a_{\rm d}$ and $v_{\rm therm}$ are the gas density, the dust mass density, the dust size and the thermal velocity($=\sqrt{8/\pi}a$) respectively.
 $n_{\rm d}$ and \red{$\Delta V$}  denote the number density of the dust($=\rho_{\rm d}/m_{\rm d}$) and the relative velocity between the dust and the gas.
 $m_{\rm d}=(4/3)\pi \rho_{\rm mat} a_{\rm d}^{3}$ is the mass of the dust and $\rho_{\rm mat}$ is the internal dust density.
We assume the Epstein drag law \citep{1924PhRv...23..710E}. The outflow velocity becomes significantly larger than the sound velocity. Therefore, equation (\ref{K}) include the correction due to the super sonic velocity in the coefficient $K$ \citep{2012MNRAS.420.2365L}. 
\red{However, for a conservative estimate, we set $\Delta V=0$ in this study.}
\subsubsection{\red{Dust Lifting Criterion}}
\red{
\red{In this study, we examine whether the outflow can lift up the dust grains by considering the dynamics of the dust grains in the poloidal direction. Under the terminal velocity approximation, the poloidal dust velocity $v_{d,p}$ is expressed as}}
\begin{equation}
\label{dust-pol1}
    \red{v_{d,p}=v_{p}+\Delta v,}
\end{equation}
where,
\begin{align}
    \textcolor{black}{\Delta v =} \textcolor{black}{-t_{\rm stop}\bigg(\frac{GM_{*}}{(r^{2}+z^{2})^{3/2}}(r\cos\alpha + z\sin\alpha) 
    - \frac{v_{\phi}^{2}}{r}\cos\alpha\bigg)}. \label{dust-pol2}
\end{align}
\red{$\Delta v$ represents the relative velocity between dust and gas along the magnetic field line (i.e., the poloidal relative velocity).} \red{$\alpha$ represents the angle between the magnetic field line and the horizontal plane. 
From the equation (\ref{dust-pol1}), the dust grains can be \red{lifted up} by the outflow if $|v_{p}|>|\Delta v|$, otherwise, they settle into the disk.}

\subsubsection{\red{Stokes Number for Dust Grains}}
\red{The Stokes number St is an indicator of how strongly gas and dust are coupled. If the Stokes number is smaller than unity, the terminal velocity approximation is valid. On the other hand, if the Stokes number is greater than unity, the terminal velocity approximation is invalid.}
 
 To calculate the Stokes number, we need to specify the dynamical timescales. In this paper, we consider following two timescales,
\begin{equation}
    t_{\rm dyn1}=\frac{1}{\Omega_{\rm g}}, \quad  \Omega_{\rm g}=\frac{v_{\phi}}{r},
\label{24}
\end{equation}
\begin{equation}
    t_{\rm dyn2}=\frac{z}{v_{p}}.
\label{25}
\end{equation}

Equation (\ref{24}) assumes that the dynamical time scale is the rotational period of the outflow, and equation (\ref{25}) assumes that the dynamical timescale is the timescale for the dust to reach a certain vertical height $z$ with a poloidal velocity $v_{p}$. When we consider the dust dynamics in a disk, $t_{\rm dyn}=1/\Omega_{\rm g}$ is often assumed. However, even if the dust can no longer follow the rotational gas motion, there is a possibility that the dust will continue to move upwards if it follows the gas-driven updraft. Therefore, we consider the dynamical timescales of both toroidal and poloidal motion in this paper. 

\subsubsection{\red{The Criteria that Determine the Fate of the Dust Grains}}
\red{In this paper, using the following conditions, we consider whether the dust grains are lifted by the outflow and escape into interstellar space.}

\red{First, we define the “escape point”, $z_{\rm esc}$. The escape point is the point at which the dust's terminal velocity exceeds the escape velocity $v_{\rm esc}$, where $v_{\rm esc}$ is defined as,}
\begin{equation}
\red{
    v_{\rm esc}=\sqrt{\frac{2GM_{*}}{R}},\quad R=\sqrt{r^{2}+z^{2}}.}
\end{equation}

\red{If the dust poloidal terminal velocity and the dust Stokes number satisfy $\St<1$ and $v_{d, p}> 0$ (or equivalently $|\Delta v|< |v_p|$ ), the dust velocity is at the terminal velocity, and it is thought that the dust grains continue to rise due to the outflow. Furthermore, if these conditions are met up to the escape point (i.e., in $z<z_{\rm esc}$), it can be regarded that the dust grains will escape into interstellar space.
Therefore, we define the conditions of $|\Delta v|< |v_p| $ and $\rm St<1$ in $z<z_{\rm esc}$ as the "dust escape condition" for a given dust size. On the other hand, if either of these conditions is not met, the dust will be unable to be lifted up or the terminal velocity approximation breaks down. In such cases, the dust motion may be complex, but the dust grains likely reaccrete onto the disk. Therefore, when the dust escape condition is not met, we express this as the dust settling case.} \red{Figure \ref{model1} provides a detailed illustration of the fate of the dust grains.}

 \subsection{Disk Model}
In this paper, we consider a gravitationally unstable heavy disk as a disk model (i.e., a disk with the Toomre's $Q$ parameter \citep{1964ApJ...139.1217T} is unity).
The Toomre's $Q$ parameter is given as,
\begin{equation}
    Q=\frac{c_{s}\Omega_{\rm disk}}{\pi G \Sigma_{\rm g}}.
\end{equation}
$c_{\rm s}$, $\Sigma_{\rm g}$ and $\Omega_{\rm disk}$ are the sound speed, the gas surface density, and the disk rotation frequency, respectively. 
The disk temperature is  assumed to be,
\begin{equation}
    T=\red{150}\bigg(\frac{r}{\rm 1AU}\bigg)^{-1/2}\rm K.
\end{equation}
The sound speed is given as, 
\begin{equation}
    c_{\rm s}=\bigg(\frac{k_{\rm B}T}{\mu m_{\rm p}} \bigg)^{1/2}= 7.3 \times10^{4}\bigg(\frac{r}{1\mathrm{AU}}\bigg)^{-1/4} \rm cm \, s^{-1},
\end{equation}
where $k_{\rm B}$, $\mu$ and $m_{\rm p}$ denote the Boltzmann constant, the mean molecular weight of the gas, and the proton mass respectively. 

\par
In our fiducial model, we assume the disk size to be 20AU, the effective height of the wind launching point to be $H_{\rm B}=0.1r$, the mass of the central protostar to be $M=0.3M_{\odot}$, and the mass ejection rate of the outflow to be $\dot{M}=10^{-6}\,M_{\odot}\,\rm yr^{-1}$.
Then the temperature and the density at the wind base are $T_0 = 33 \rm K$ and 
$\rho_{0}=2.8\times10^{-12} \rm g\,cm^{-3}$, respectively (see \autoref{para-fid} for details).

\section{RESULTS} \label{sec:floats}
\subsection{Results of the Fiducial Model}
\label{fid_ex}
\subsubsection{Velocity Structure}
\begin{figure}
    \centering
    \includegraphics[scale=0.45]{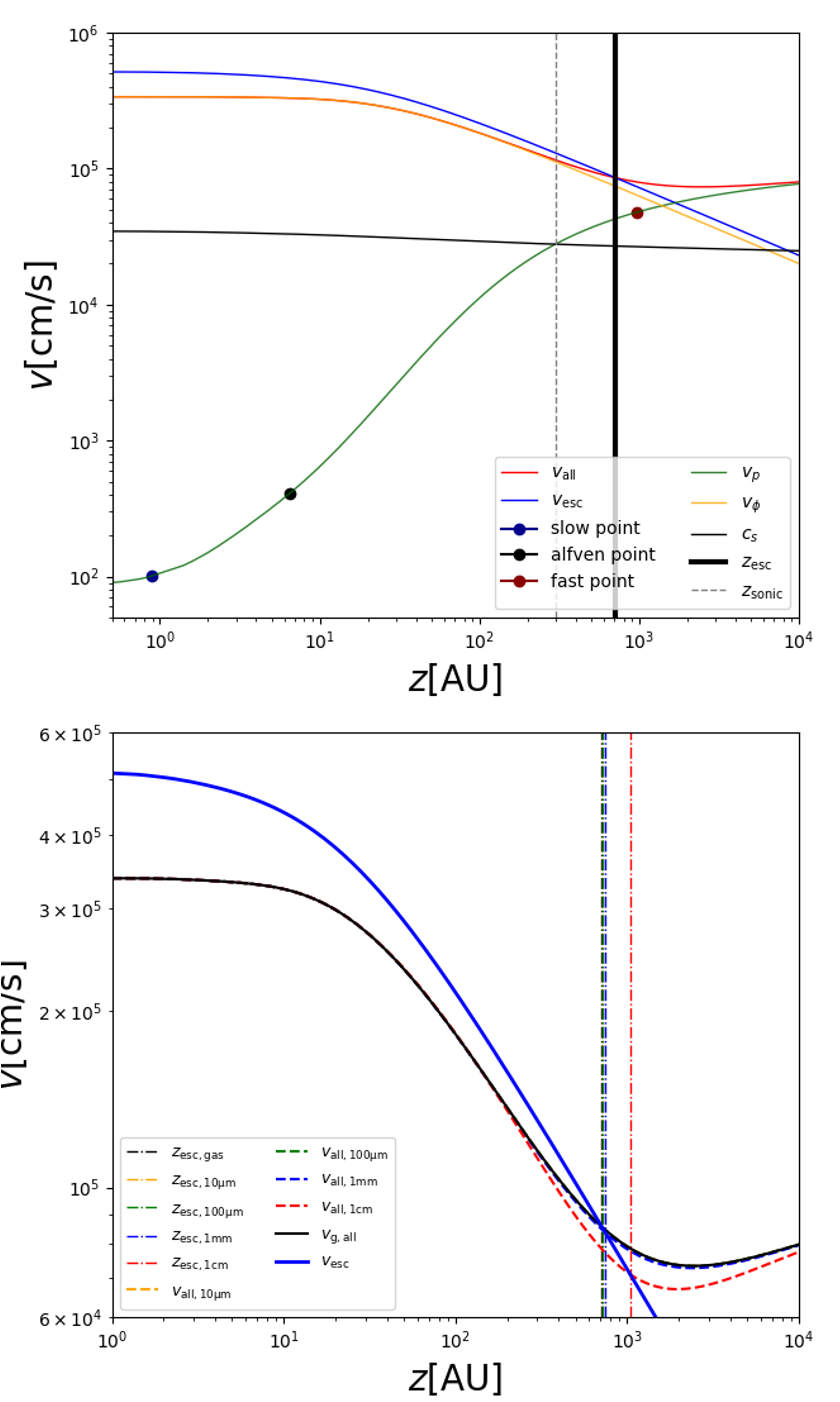}
    \caption{Top: The velocity structure of the outflow as a function of  $z$. The green, orange, and red lines show the poloidal velocity $v_{ p}$, the toroidal velocity $v_{\rm \phi}$, and the total velocity ($\sqrt{v_{ p}^2+v^2_{\rm \phi}}$) of the outflow, respectively. The black and blue lines show the sound velocity $c_{\rm s}$  and the escape velocity $v_{\rm esc}$, respectively. Vertical thick black solid line and vertical gray dashed line show the hight where $\sqrt{v_{ p}^2+v^2_{\rm \phi}}=v_{\rm esc}$ and $v_{ p}=c_{\rm s}$ (sonic point). Bottom: A comparison between the total gas velocity and the total dust velocity ($\sqrt{v_{ d,p}^2+v^2_{\rm \phi}}$). The orange, green, blue, red lines show the total dust velocity of 10 $\rm \mu m$, 100 $\rm \mu m$, 1 mm, 1 cm, respectively. The vertical dashed-dotted lines show the escape point of different dust sizes.}
    \label{fidv}
\end{figure}

The top panel of the figure \ref{fidv} shows the wind structure of our fiducial model. 
The poloidal velocity $v_{p}$ (green solid line) monotonically increases from the wind base to the upper layers and eventually reaches the terminal velocity. 
On the other hand, the toroidal velocity $v_{\phi}$ (orange solid line) monotonically decreases and obeys $v_{\phi}\propto z^{-0.5} (\propto r^{-1})$ after passing through the Alfv$\Acute{\rm e}$n point which shows the conservation of the angular momentum.  
The total velocity of the outflow (red solid line) exceeds the escape velocity (blue solid line) at the height of $\sim 700$AU (vertical black solid line). This line shows the border between the region where the gas is gravitationally unbound and the gas is gravitationally bound. Note that the poloidal velocity of the outflow exceeds the sound speed at a considerably high position compared to the wind base (vertical gray dashed line). In the region below this, the gas is also being accelerated by the magnetic field. However, the poloidal velocity is small, so it may not be regarded as an outflow in observational or simulation analysis.  \par
\red{The bottom panel of the figure \ref{fidv} shows the comparison between the dust (total) terminal velocity for each dust grain size with the gas total velocity. From this figure, we can see that the escape point of the dust grains varies according to the dust size. This is because $t_{\rm stop}$
becomes larger for larger dust size resulting in the smaller poloidal dust velocity $v_{d,p}$.} \red{However, except for the 1 cm sized dust grains, the escape point $z_{\rm esc}$ of the gas and the dust are nearly the same.}

\begin{figure*}[t]
    \centering
    \includegraphics[scale=0.57]{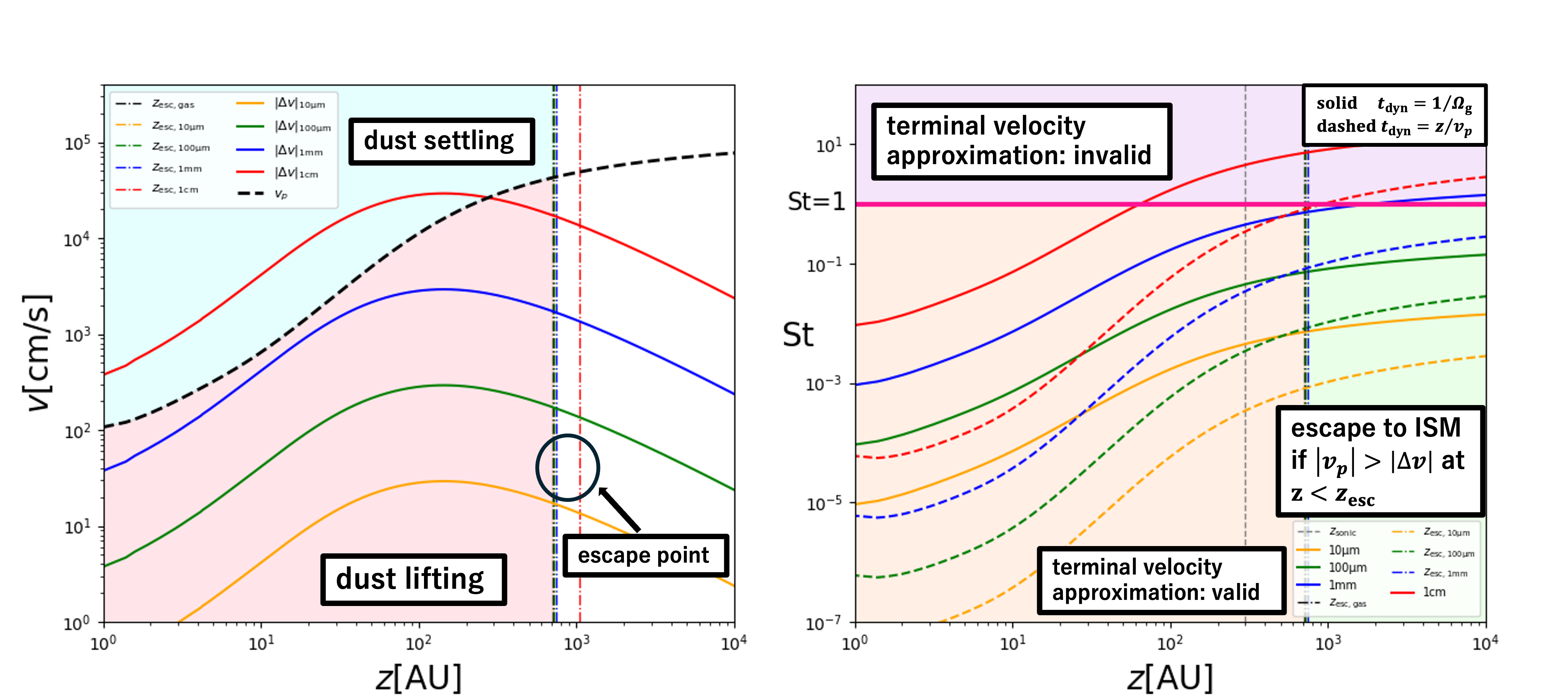}
    \caption{The results of the fiducial model (with the parameters $\dot{M} =10^{-6}\,M_{\odot}\, \rm yr^{-1}$, $M_{*}=\red{0.3}\,M_{\odot}$, $r_{\rm disk}=\red{20}\mathrm{AU}$, $\rho_{\rm mat}=1\,\rm g\,cm^{-3}$, and $H_{\rm B}=0.1r$). \red{Left: A comparison between the poloidal relative velocity $|\Delta v|$ (solid line) and the gas poloidal velocity $v_{p}$ (black dashed line). The orange, green, blue, and red lines show $\Delta v$ of 10 $\mu$m, 100 $\mu$m, 1 mm, and 1 cm, respectively. The vertical dashed-dotted line shows the escape point of different dust sizes and gas.} Right: The Stokes number of dust as a function of $z$. The orange, green, blue, red lines show the Stokes number of 10 $\mu$m, 100 $\mu$m, 1 mm, and 1 cm, respectively. $t_{\rm dyn1}$ and $\rm t_{\rm dyn2}$ are shown by solid and dashed lines respectively. The pink horizontal line shows the position where St equals unity.} 
    \label{fig:enter-label}
    \label{fid}
\end{figure*}

\subsubsection{\red{Dust Dynamics in the Outflow: Dust Lifting and Escape}}

\red{ The left panel of the figure \ref{fid} shows the poloidal relative velocity $|\Delta v|$ and the gas poloidal velocity $|v_{p}|$ (black dashed line). 
The colors of the lines represent different dust sizes. The vertical dashed-dotted lines represent the escape point with different dust sizes. 
The red-hatched area is the region where $|v_{p}|>|\Delta v|$ in $z<z_{\rm esc}$ and the dust grains continue to rise with their terminal velocity. The blue-hatched area is the region where $|v_{p}|<|\Delta v|$ in $z<z_{\rm esc}$. Once a line enters this region, it shows that the dust grains cannot be lifted by the outflow and may settle into the disk. }

\red{By comparing the black dashed line and color lines, we can conclude that the dust grains of $\lesssim$ 1 mm can be lifted by the magnetically driven outflow with our fiducial parameter if the terminal velocity approximation is valid.}

\begin{table}[t]
    \centering
    \caption{\textit{Parameters and Values of Fiducial Model}} 
    \label{para-fid}
    \hspace{-1cm}
    \scalebox{0.8}{ 
    \begin{tabular}{lc}
        \hline
        \textbf{Parameter} & \textbf{Value} \\
        \hline
        $\dot{M}$ (mass ejection rate of the outflow) & $10^{-6} \,M_{\odot}\,\rm yr^{-1}$ \\
        \red{$M_{*}$} (mass of the central protostar)  & $\red{0.3}\,M_{\odot}$ \\
        $r_{\rm disk}$ (disk radius)  & $\red{20}\,\mathrm{AU}$ \\
        $T_{0}$ (temperature)  & $\red{33}\,\mathrm{K}$ \\
        $\rho_{\rm mat}$ (internal dust density)  & $1\,\rm g\,cm^{-3}$ \\
         $H_{B}$ (effective height of wind launching point)  & $0.1 r$ \\
        $\rho_{0}$ (density)  & $2.8 \times 10^{-12}\,\rm g\,cm^{-3}$\\
        $\Theta$ (inclination angle)  & $\red{53}^{\circ}$ \\
        $B_{p}$ (poloidal magnetic field) & $\red{1.3 \times 10^{-3}} \mathrm{G}$ \\
        \hline
    \end{tabular}
    }
\end{table}

In the right panel \red{of figure \ref{fid}}, the \red{orange}-hatched area is the region where $\St<1$ and \red{$z<z_{\rm esc}$}. In this region, the gas and the dust grains are coupled to each other while moving, \red{so the terminal velocity approximation is valid in this area} and both gas and dust are gravitationally bound to the central protostar potential. The green-hatched area is the region where $\St<1$ and \red{$z>z_{\rm esc}$}. In this region, the gas and the dust grains are coupled and neither the gas nor the dust is gravitationally bound. The \red{purple}-hatched area is the region where $\St>1$ and \red{$z<z_{\rm esc}$}. \red{In this region, the terminal velocity approximation is invalid before the dust velocity exceeds the escape velocity. If a line enters this region, our escape condition is not met, so we consider that the dust grains cannot escape into interstellar space due to the outflow. }
\begin{figure*}[t]
         \centering
        \includegraphics[scale=0.35]
             {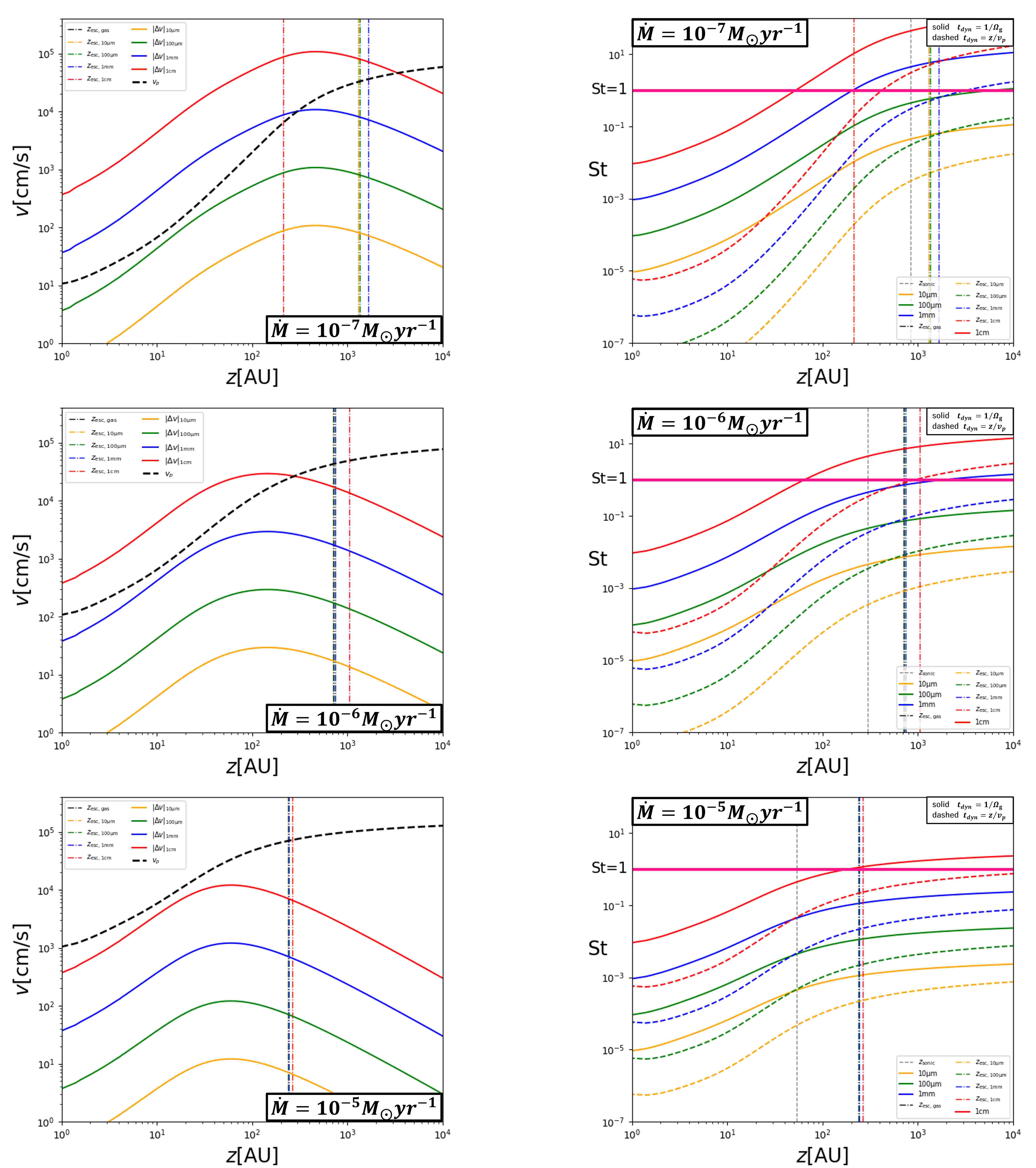}
    \caption{Left: A comparison between the poloidal relative velocity $|\Delta v|$ and the gas poloidal velocity $|v_{p}|$ (black dashed line). Right: The Stokes number of the dust as a function of $z$. The mass ejection rate of the outflow $\dot{M}$ are $10^{-7}$ (top), $10^{-6}$ (middle), and $10^{-5}~M_{\odot} {\rm yr^{-1}}$ (bottom), respectively. The poloidal magnetic fields are $\red{1.3}\times10^{-3}\, \mathrm{G}$ (top), $\red{9.5}\times10^{-4} \,\mathrm{G}$ (middle) and $\red{6.8}\times10^{-4}\, \mathrm{G}$ (bottom) respectively. The other parameters are the same as the fiducial model  (table 1).}
    \label{mdot1}
\end{figure*}

\begin{figure*}[t]
         \centering
         
    \includegraphics[scale=0.35]{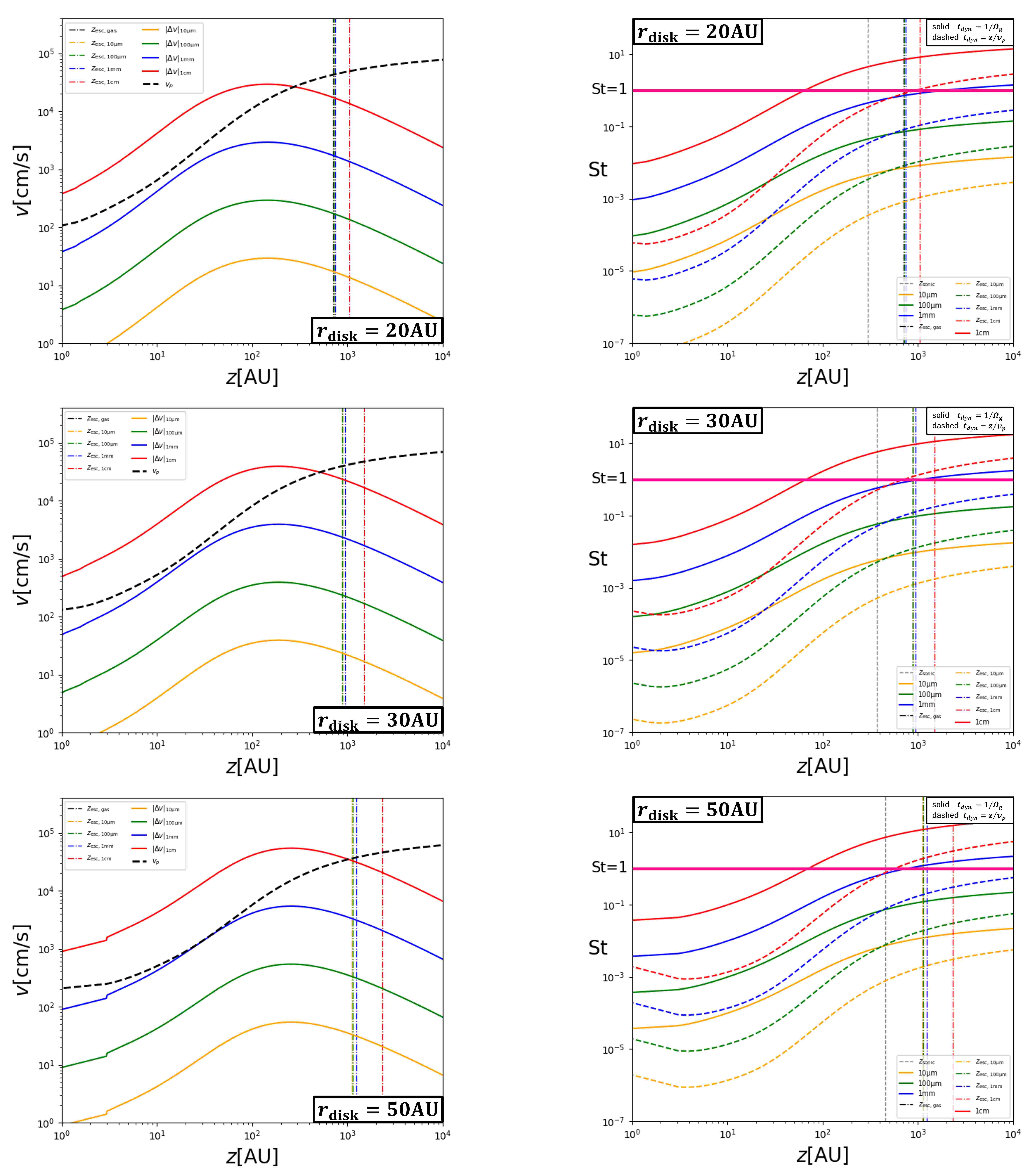}
    \caption{Left: A comparison between the poloidal relative velocity $|\Delta v|$ and the gas poloidal velocity $|v_{p}|$ (black dashed line). Right: The Stokes number of the dust as a function of $z$. The disk size are  $r_{\rm disk}=20$ (top), $r_{\rm disk}=30$ (middle), and $r_{\rm disk}= 50\mathrm{AU}$ (bottom), respectively. The poloidal magnetic fields are $\red{1.3}\times10^{-3}\, \mathrm{G}$ (top), $\red{9.5}\times10^{-4} \,\mathrm{G}$ (middle) and $\red{6.8}\times10^{-4}\, \mathrm{G}$ (bottom) respectively. The other parameters are the same as the fiducial model  (table 1).}
    \label{disk1}
\end{figure*}

\begin{figure*}[p]
         \centering
    \includegraphics[scale=0.35]{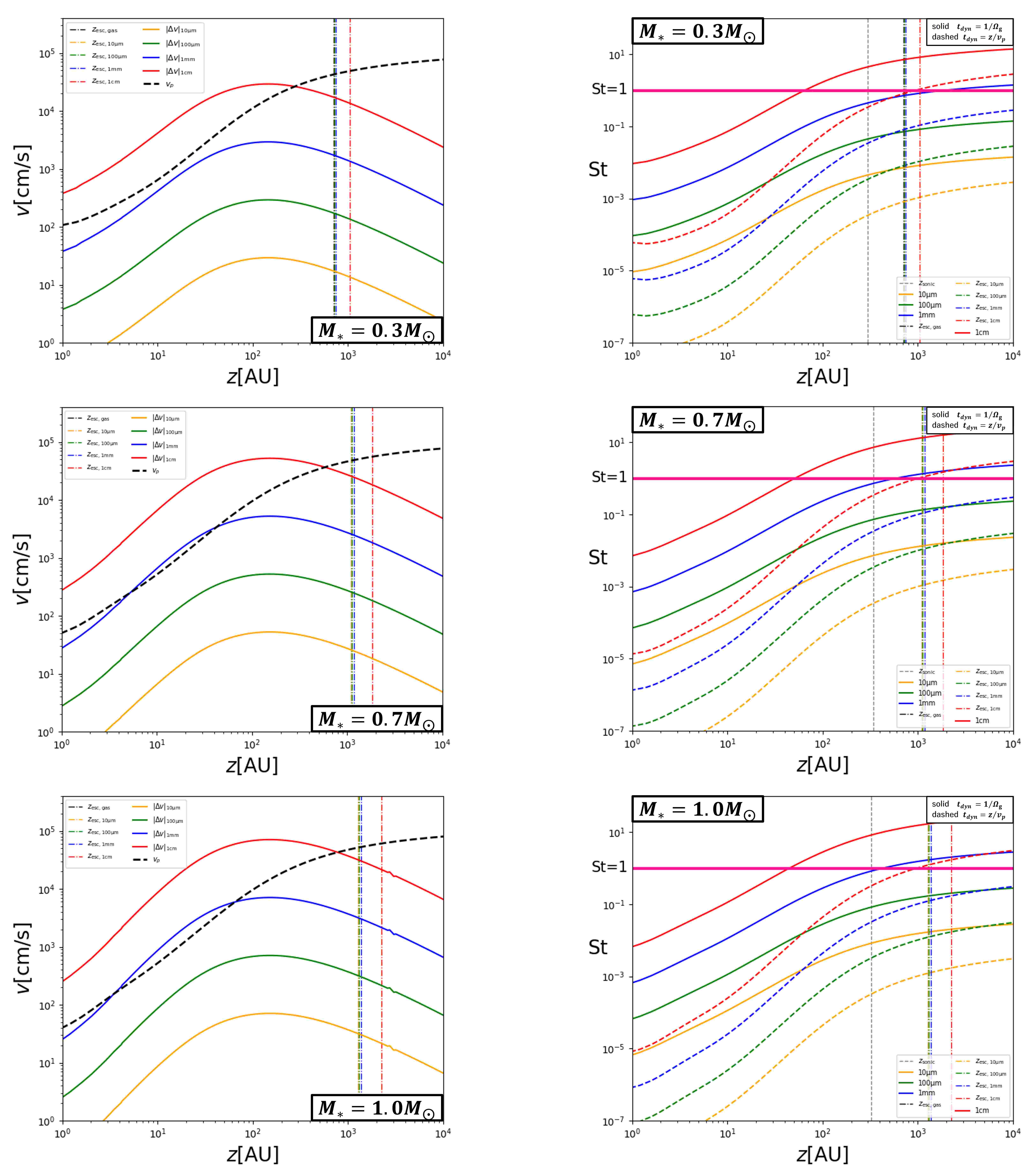}
    \caption{Left: A comparison between the poloidal relative velocity $|\Delta v|$ and the gas poloidal velocity $|v_{p}|$ (black dashed line). Right: The Stokes number of the dust as a function of $z$. The Mass of central protostar are  $M_{*}=\red{0.3} M_\odot$ (top), $M_{*}=\red{0.7} M_\odot$ (middle), and $M_{*}=\red{1.0} M_\odot$ (bottom), respectively. The poloidal magnetic fields are $\red{1.3}\times10^{-3}\, \mathrm{G}$ (top), $\red{1.2}\times10^{-3} \,\mathrm{G}$ (middle) and $\red{1.2}\times10^{-3}\, \mathrm{G}$ (bottom) respectively. The other parameters are the same as the fiducial model  (table 1).}
    \label{mass1}
\end{figure*}

The colors of the lines of the right panel represent different dust sizes, and the solid and dashed lines represent the $\St$ of dust in the outflow with $t_{\rm dyn1}$ and $t_{\rm dyn2}$, respectively. \red{The vertical dashed-dotted
lines represent the escape point with different dust sizes
and gas.}

\red{ 
The blue line corresponding to the dust grains with size of $1$ mm enters the green area from the \red{orange} area even with the conservative dynamical timescale ($t_{\rm dyn1}$). }
\red{Therefore,  the terminal velocity approximation is valid even for the 1 mm sized dust grains. 
}

\red{From the two panels of figure \ref{fid}, we conclude that the dust grains of $\sim 1$ mm in size can escape into interstellar space by the magnetically driven outflow with our fiducial parameter.
Note that the blue line approaches $\St=1$ near the escape point in the right panel, and the application of the terminal velocity approximation becomes delicate there. Note, however, that $\Delta v$ is already more than one order of magnitude smaller than the gas velocity there. Therefore, the slight deviation from the terminal velocity is not expected to affect the conclusion that the $1$ mm sized dust grains can escape.
}

\subsection{\red{Parameter Study of Dust Dynamics}}
\label{lift}

\subsubsection{\red{Dependence on Mass Ejection Rate of the Outflow}}
\red{The left panel of figure \ref{mdot1} shows the poloidal relative velocity $|\Delta v|$ and the gas poloidal velocity $|v_{p}|$ (black dashed line) with the different mass ejection rate of the outflow $\dot{M}$. The right panel of figure \ref{mdot1} shows St with the different mass ejection rate of the outflow $\dot{M}$.} Top, middle, and bottom panels show the results with $\dot{M}=10^{-7} M_{\odot}\rm yr^{-1}$, $\dot{M}=10^{-6} M_{\odot} \rm yr^{-1}$ (fiducial model), $\dot{M}=10^{-5} M_{\odot}\rm yr^{-1}$, respectively. 

\red{As the mass ejection rate increases, the outflow entrains larger-sized dust grains. 
When the mass ejection rate of the outflow is $\dot{M}=10^{-5} M_{\odot}\rm yr^{-1}$, $|\Delta v|$ of 1 cm sized dust grains (red solid line) becomes smaller than the gas poloidal velocity meaning that even the 1 cm sized dust grains can be lifted up by the outflow. 
Note that St becomes $\St\sim 1$ around the escape point for the 1 cm sized grains as shown in the right panel of the figure \ref{mdot1}. Thus, the validity of the terminal velocity approximation is debatable there.
However, even if the dust velocity does not completely converge to the terminal velocity,  it is expected that the relative velocity may be much smaller compared to the gas velocity around the escape point, as seen by the comparison of the terminal velocity (red solid line) and the gas velocity (black dashed line).
Therefore, the conclusion that the 1 cm sized dust grains will be lifted up would not change.}

\red{The maximum dust size which can be lifted up by the outflow strongly depends on $\dot{M}$.
When the mass ejection rate of the outflow is $\dot{M}=10^{-7} M_{\odot}\rm yr^{-1}$, the maximum size of dust grains becomes $\lesssim 100 \rm\mu m$.}
\red{An increase in the mass ejection rate of the outflow causes an increase in the gas poloidal velocity $v_{p}$ (because the density of the wind base (i.e., disk) is fixed). On the other hand, $|\Delta v|$ does not depend on the mass ejection rate and is constant (see the left panel of figure \ref{mdot1}). This is the reason why an increase in the outflow rate increases the maximum size of dust grains that can be lifted up by the outflow.}

\subsubsection{\red{Dependence on Disk Size}}
\red{The left panel of figure \ref{disk1} shows the poloidal relative velocity $|\Delta v|$ and the gas poloidal velocity $|v_{p}|$ (black dashed line) with different disk sizes $r_{\rm disk}$. The right panel of figure \ref{disk1} shows $\St$ with different disk size $r_{\rm disk}$. The top, middle and bottom panels show the results with $r_{\rm disk}=20 {\rm AU}$ (fiducial model), $r_{\rm disk}=30 {\rm AU}$, $r_{\rm disk}=50 {\rm AU}$, respectively. } 

\red{The dust size that is lifted by outflow decreases with an increase in the disk size. The upper left panel ($r_{\rm disk}=20\rm AU$) shows that the blue line (1 mm sized dust grains) does not contact the black dashed line, and the poloidal velocity of the dust grains $v_{d,p}$ is positive, so the dust grains continue to rise. 
On the other hand, the lower left panel ($r_{\rm disk}=50\rm AU$) shows that the blue line contacts the black dashed line, and the poloidal velocity of the dust becomes less than 0 around the point, so the dust grains stop rising there. }

\red{However, the impact of the disk size on the maximum dust size that is lifted is not strong. This is because the increase in the gas poloidal velocity partially cancels out the increase in $|\Delta v|$ caused by the decrease in the density in the outflow.}

\red{The right panels also show that the dependence of St on the disk size is not strong. As discussed in the
previous section, in the region where St is close to 1 (for example, $z \sim 10^3$ AU), the value of $|\Delta v|$ is much smaller than $|v_p|$. On the other hand, in the region where the value of $|\Delta v|$ is close to $|v_p|$ (e.g. $z \sim 20$ AU), St is sufficiently small even for the dust grains of 1 mm in size, and the terminal velocity approximation is valid there. Therefore, it is expected that the dust grains of around 1 mm in size are lifted up even for a 50 AU sized disk with our fiducial parameters.}

\subsubsection{\red{Dependence on Mass of the Central Protostar}}
\red{The left panel of figure \ref{mass1} shows the poloidal relative velocity $|\Delta v|$ and the gas poloidal velocity $|v_{p}|$ (black dashed line) with the different mass of the central protostar $M_{*}$. The right panel of figure \ref{mass1} shows the $\St$ with the different mass of the central protostar \red{$M_{*}$}. Top, middle, and bottom panels show the results with $M_{*}=0.3 M_\odot$ (fiducial model), $M_{*}=0.7 M_\odot$,  $M_{*}=1.0 M_\odot$, respectively.} \par

\red{ The size of the dust grains lifted by the outflow decreases as the mass of the central protostar increases. In the upper left panel ($M_{*}=0.3 M_\odot$), the blue line (1 mm sized dust grains) lies below the black dashed line. In the middle left ($M_{*}=0.7 M_\odot$) and lower left ($M_{*}=1.0 M_\odot$) panels, on the other hand, there is the region (at $z\sim 20$ AU) where the blue line lies above the black dashed line. In this region, the dust grains do not follow the upward gas flow and settle.} 

\red{The cause of the central star mass affecting the maximum dust size is mainly due to the decrease in the gas poloidal velocity due to the increase in disk density (because we consider a self-gravitationally unstable disk).}

\red{The right panels show that several $100 ~\rm\mu m$ sized dust grains satisfy $\St<1$, and the terminal velocity approximation is valid.
Therefore, we conclude that it is expected that the dust grains with the size of several 100 $ \rm \mu m$ are lifted up by the outflow but the dust grains with the size of 1 mm are not lifted up for the protostar with the mass of $\sim 1 M_\odot$ with our fiducial parameters.}

\begin{figure*}[t]
         \centering
         
    \includegraphics[scale=0.35]{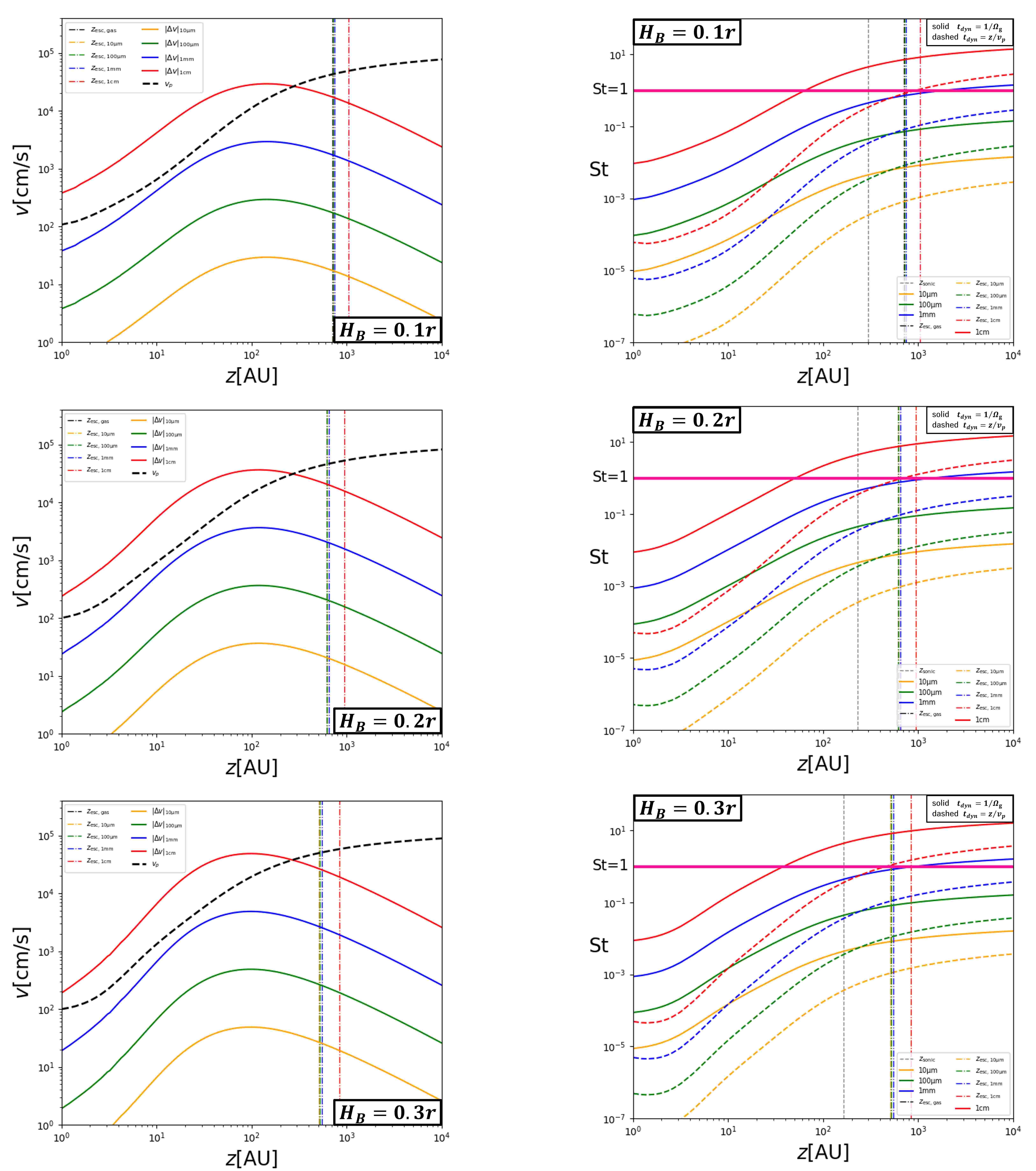}
    \caption{Left: A comparison between the poloidal relative velocity $|\Delta v|$ and the gas poloidal velocity $|v_{p}|$ (black dashed line). Right: The Stokes number of the dust as a function of $z$. The values of $H_{\rm B}$  are $H_{\rm B}=0.1r$ (top), $H_{\rm B}=0.2r$ (middle), and $H_{\rm B}=0.3r$ (bottom), respectively. The poloidal magnetic fields are $\red{1.3}\times10^{-3}\, \mathrm{G}$ (top), $\red{1.6}\times10^{-3} \,\mathrm{G}$ (middle) and $\red{2.1}\times10^{-3}\, \mathrm{G}$ (bottom) respectively. The other parameters are the same as the fiducial model (table 1).}
    \label{HB1}
\end{figure*}

\subsubsection{\red{Dependence on the Launch Point of the Outflow}}
\label{z0_dependence}

\red{As \citet{2016ApJ...818..152B} pointed out, the launching point of outflows may be the disk surface because the gas and magnetic field may decouple around the midplane.
Therefore, it is important to investigate whether our results are affected by the height of the launch point.}
As we discussed in section \ref{sec:style},
$H_{\mathrm{B}}$  can be regarded as the parameter that controls the height of the launching point of the outflow in our outflow model.  
Therefore, we investigate how the maximum dust size depends on the launching point by changing $H_{\mathrm{B}}$.

\red{The left panel of figure \ref{HB1} shows the poloidal relative velocity $|\Delta v|$ and the gas poloidal velocity $|v_{p}|$ (black dashed line) with the different value of $H_{\rm B}$. The right panel of figure \ref{HB1} shows the St with the different value of $H_{\rm B}$. Top, middle and bottom panels show the results with $H_{\rm B}=0.1r$ (fiducial model), $H_{\rm B}=0.2r$, and $H_{\rm B}=0.3r$, respectively.  } 

\red{As the figure shows, $H_{\rm B}$ does not affect the velocity of the gas and dust grains, or the St of the dust grains. Therefore, we conclude that the uncertainty of the launching point of the outflow does not affect our results.}

\subsubsection{\red{Dependence on the Inclination Angle between the Disk Surface and Magnetic Field Line}}
\label{sec_inc}

\begin{figure}
    \centering
    \includegraphics[scale=0.47]{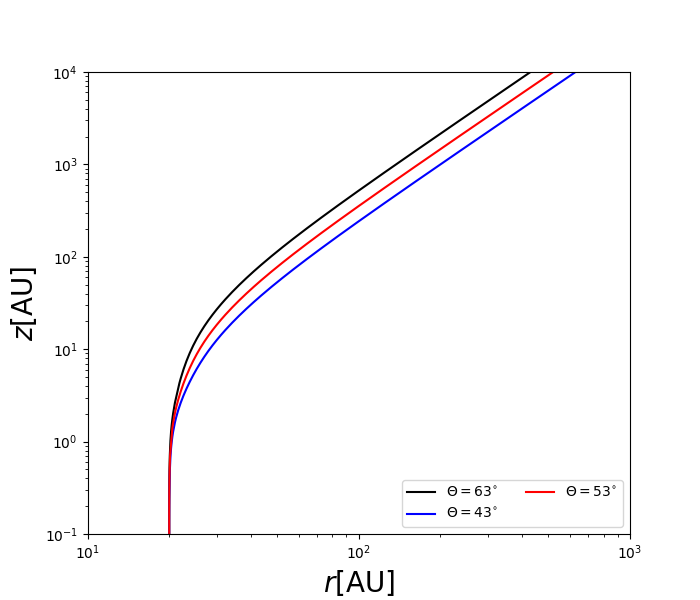}
    \caption{The configuration of the magnetic field lines with different inclination angle between the disk surface and the field line $\Theta$. The blue, red, and black lines represent $\Theta=43^\circ$, $\Theta=53^\circ$, and $\Theta=63^\circ$. respectively. }
    \label{magline}
\end{figure}

\begin{figure*}[t]
    \centering
    \includegraphics[scale=0.58]{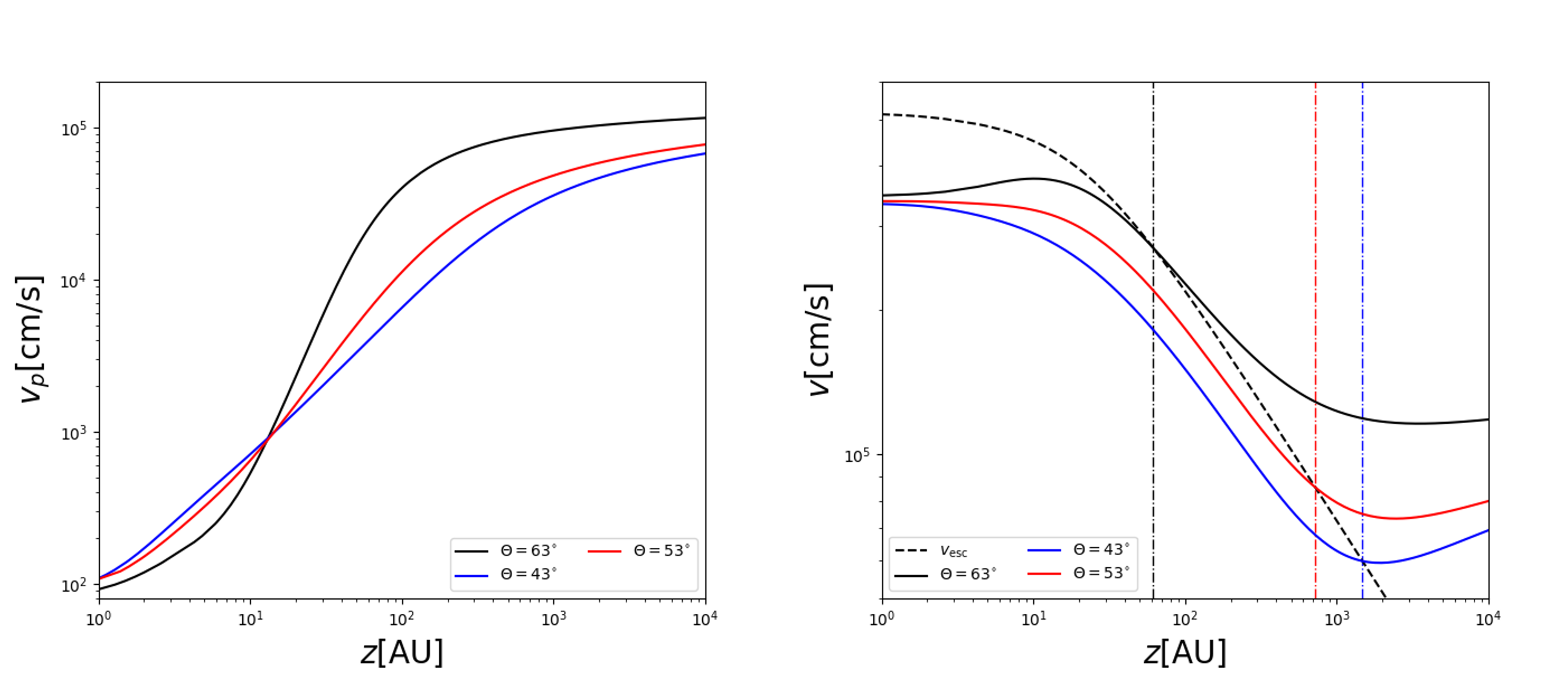}
    \caption{Velocity structure of the outflow as a function of $z$ when the inclination angle between the disk surface and the magnetic field line $\Theta$ is changed. Left: Poloidal velocity as a function of $z$. Right: Total velocity of the outflow as a function of $z$. The blue, red and black lines show the results with $\Theta=43^{\circ}$, $\Theta=53^{\circ}$, and $\Theta=63^{\circ}$, respectively. The vertical dashed-dotted lines show the escape point of different $\Theta$.   }
    \label{inc2}
\end{figure*}

As first pointed out by \citet{1982MNRAS.199..883B}, the structure of the outflow strongly depends on the inclination of the magnetic field $\Theta$. Of particular importance in the present study is the fact that, as \citet{1997ApJ...474..362K} showed, the outflow velocity changes even with the same magnetic energy.
\red{The left panel of the figure \ref{inc2} shows the poloidal gas velocity $v_{p}$ for the different inclination angle $\Theta$ (the figure \ref{magline} shows the configuration of the magnetic field lines with different $\Theta$). This difference causes the change of the escape point.}

\red{The right panel of the figure \ref{inc2} shows the total gas velocity for the different inclination angles. When the inclination angle $\Theta$ is $\Theta=43^\circ$ (\red{blue} solid line), the outflow exceeds the escape velocity (black dashed line) at $z\sim\red{2}000 \mathrm{AU}$. On the other hand, with $\Theta=63^\circ$ (\red{black} solid line), the height at which the outflow velocity exceeds the escape velocity is at $z\sim\red{6}0\mathrm{AU}$. This shows that the height at which the outflow velocity exceeds the escape velocity strongly depends on the inclination angle between the disk surface and the magnetic field line. }

\begin{figure*}[t]
         \centering
       
    \includegraphics[scale=0.35]{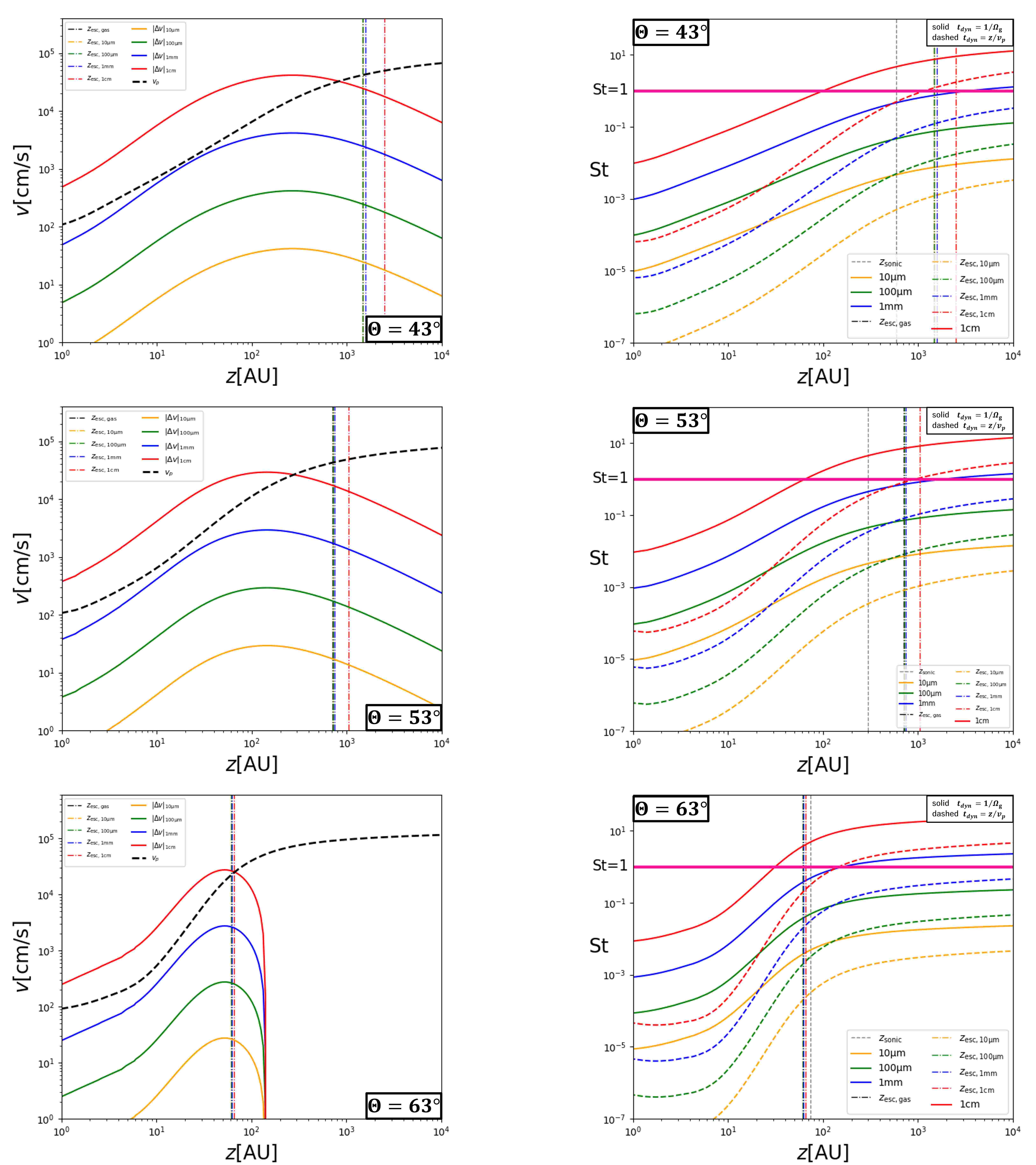}
    \caption{Left: A comparison between the poloidal relative velocity $|\Delta v|$ and the gas poloidal velocity $|v_{p}|$ (black dashed line). We only plot $|\Delta v|$ when it is negative.
    Right: The Stokes number of the dust as a function of $z$. Inclination angle between the disk surface and magnetic field line are $\Theta=43^{\circ}$ ($r_{\mathrm{N}}/r_{0}=100$) (top), $\Theta=53^{\circ}$ ($r_{\mathrm{N}}/r_{0}=2.5$)  (middle), and $\Theta=63^{\circ}$ ($r_{\mathrm{N}}/r_{0}=1.17$) (bottom), respectively. The poloidal magnetic fields are $\red{1.2}\times10^{-3}\, \mathrm{G}$ (top), $\red{1.3}\times10^{-3} \,\mathrm{G}$ (middle) and $\red{1.5}\times10^{-3}\, \mathrm{G}$ (bottom) respectively. The other parameters are the same as the fiducial model (table 1).}
    \label{incl1}
\end{figure*}

\red{The left panel of figure \ref{incl1} shows the poloidal relative velocity $|\Delta v|$ and the gas poloidal velocity $|v_{p}|$ (black dashed line) with the different inclination angle $\Theta$ and note that we only plot $|\Delta v|$ when it is negative. The right panel of figure \ref{incl1} shows the St with the different inclination angle $\Theta$. Top, middle and bottom panels show the results with $\Theta=43^\circ$ ($r_{\mathrm{N}}/r_{0}=100$), $\Theta=53^\circ$(fiducial model), and $\Theta=63^\circ$ ($r_{\mathrm{N}}/r_{0}=1.17$), respectively.} 

\red{The left panel of figure \ref{incl1} shows that the $|\Delta v|$ profiles are very similar for $\Theta=43^\circ$ $\Theta=53^\circ$ ($|\Delta v|$ with $\Theta=43^\circ$ case is slightly larger, though). On the other hand, the $|\Delta v|$ profiles change significantly between these and $\Theta=63^\circ$. In particular, $|\Delta v|$ decreases rapidly in all dust sizes at $z \sim 100 \rm AU$. This shows that $\Delta v>0$ there. The reason why $\Delta v>0$ is realized is that the centrifugal force becomes larger than the gravity force by the central star. Therefore, if we focus on the dependence of dust $\Delta v$ on $\Theta$, there is a major difference.
}

\red{
 However, if we focus on the maximum dust size that can be lifted, the larger/smaller relationship between the dashed and solid lines does not change with $\Theta$. 
Similarly, the right panels show that both the change in the St and the change in the escape point occur in response to $\Theta$. As a result, the condition of whether or not the St of a given dust size exceeds 1 below the escape point also does not change.
 Therefore, the conclusion that the dust grains up to $1$ mm in size can be lifted is maintained regardless of the value of $\Theta$. }

\subsection{\red{Dependence of Maximum Lifting Dust Size on Outflow Parameters}}
In this section, we investigate the maximum size of dust grains that \red{can be lifted by the outflow} ($a_{\rm d,max}$).

\begin{figure*}[t]
    \includegraphics[scale=0.75]{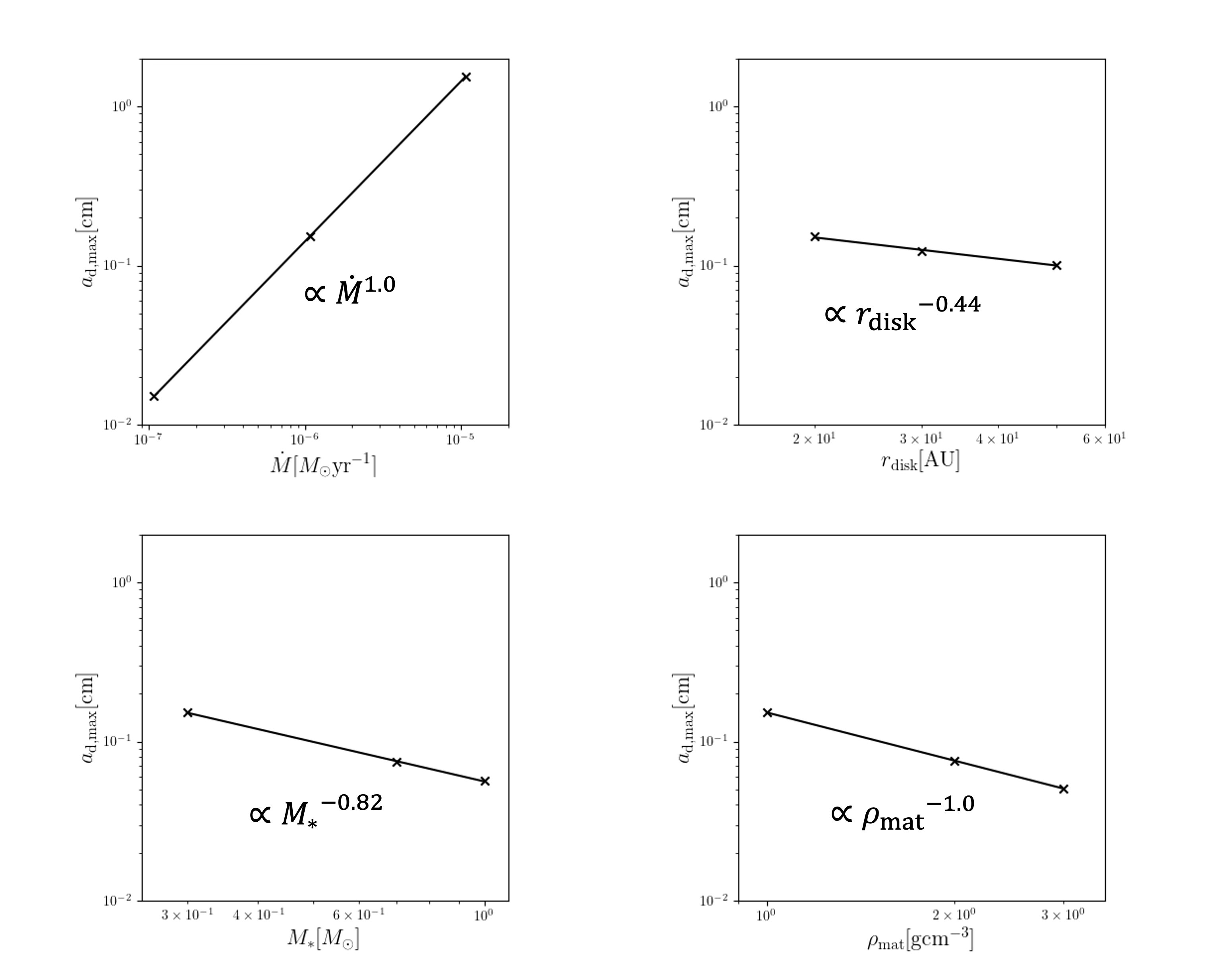}
    \centering
    
    \caption{Maximum size of the dust grains with which the dust grains are lifted by the outflow $a_{\rm d,max}$ as a function of mass ejection rate of the outflow $\dot{M}$, the disk size $r_{\rm disk}$, the mass of the central protostar $M_{*}$ and the internal dust density $\rho_{\rm mat}$.}
    \label{paradep}
\end{figure*}

Figure \ref{paradep} shows the dependence of $a_{\rm d,max}$ on the mass ejection rate of the outflow $\dot{M}$ (top left panel), the disk size $r_{\rm disk}$ (top right panel), the mass of the central protostar $M_{*}$ (bottom left panel), and the internal dust density $\rho_{\rm mat}$ (bottom right panel).
The cross points show the calculation results, and the lines show the empirical fitting by the least squares method using the points.

$a_{\rm d,max}$ depends on the mass ejection rate $\dot{M}$, as $a_{\rm d,max}\propto \dot{M}^{1.0}.$
Increasing the outflow rate corresponds to an increase in the poloidal magnetic field $B_{p}$, which in turn increases the poloidal velocity $v_{p}$. \red{On the other hand, the value of $|\Delta v|$ is almost constant (see the left panel of figure \ref{mdot1}).} As a result, the dust grains with larger size can \red{be lifted by the outflow}.

$a_{\rm d,max}$ depends on the disk size, as $a_{\rm d,max}\propto r_{\rm disk}^{-\red{0.44}}$. 
As the disk size increases, the density at the wind base decreases, \red{which in turn leads to an increase in the stopping time $t_{\rm stop}$ and an increase in the poloidal velocity $v_{p}$.} \red{The decrease in the density with the increasing radius ($\rho  \propto  \Sigma/H  \propto  \Omega_{\rm disk}^2  \propto  r^{-3}$) is faster than the increase in the poloidal velocity ($v_{p}\propto r^{1}$) under the same mass ejection rate $\dot{M}$. As a result, the maximum size of dust grains that can be lifted by the outflow decreases with the increasing disk size.}

$a_{\rm d,max}$ depends on the mass of the central protostar \red{$M_{*}$}, as $a_{\rm d,max} \propto M_{*}^{-\red{0.82}}$. An increase in the mass of the central protostar causes an increase in the density at the wind base (because we are assuming a gravitationally unstable disk), \red{resulting in a decrease in the gas poloidal velocity $v_{p}$}. The first effect makes it easier for the larger-sized dust grains to be lifted, while the last effect makes it difficult. As a result, the increase in the mass of the central protostar makes it difficult for the larger-sized dust grains to \red{be lifted by the outflow.}

$a_{\rm d,max}$ depends on the internal dust density $\rho_{\rm mat}$ as $a_{\rm d,max} \propto \rho_{\rm mat}^{-1.0}$ because the stopping time depends on the internal density as $t_{\rm stop} \propto \rho_{\rm mat}^{1.0}$. 

By summarizing the above results, we suggest the following empirical formula for the maximum dust size which can be lifted by outflow as,
\begin{equation}
\begin{split}
    a_{\rm d,max}\sim 1.5 \bigg(\frac{\rho_{\rm mat}}{1\rm g\,cm^{-3}}\bigg)^{-1.0} \bigg(\frac{\dot{M}}{10^{-6}M_{\odot}\rm yr^{-1}}\bigg)^{1.0}  \\ \bigg(\frac{r_{\rm disk}}{20\rm AU}\bigg)^{-0.44} \bigg(\frac{M}{0.3M_{\odot}}\bigg)^{-0.82} \rm mm. 
    \label{33}
\end{split}
\end{equation}
Because the dependence of $a_{\rm d, max}$ on $H_{\rm B}$ and $\Theta$ were not significant, we do not include the dependence on them in the above equations. 
Note however that these equations can be used for at least $H_{\rm B} \le 0.3r$ because the results is almost independent of the $H_{\rm B}$ (see, section \ref{z0_dependence}).

\begin{figure}
    \centering
    \includegraphics[scale=0.5]{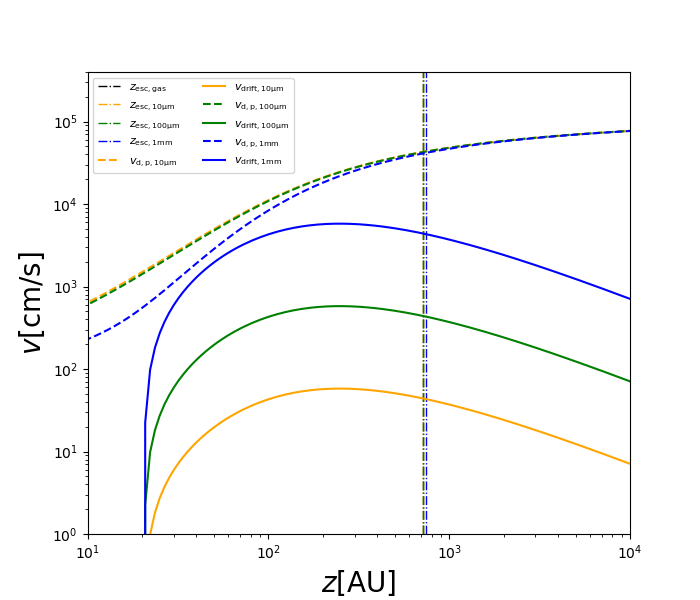}
    \caption{A comparison between the poloidal dust velocity $v_{d,p}$ (dashed) and the drift dust velocity $v_{\rm drift}$ (solid) as a function of z. The orange, green, blue lines represent 10 $\rm \mu m$, 100 $\rm \mu m$, 1 mm respectively. The vertical dashed-dotted lines show the escape point with the different dust sizes and gas.}
    \label{vd}
\end{figure}

\section{Summary \& Discussion}\label{sec:floats}

In this paper, we investigated the dust dynamics in the outflow with the analytical outflow model proposed by \cite{1997ApJ...474..362K}.
We solved one-dimensional steady and axisymmetric MHD equations to determine the structure of the outflow. Then, we analyzed the dust dynamics in the outflow focusing on the maximum dust size at which the dust grains can be \red{lifted} by the magnetically driven outflow $a_{\rm d,max}$.

We obtained the parameter dependence $a_{\rm d,max}$ on the mass ejection rate of the outflow $\dot{M}$, the disk size $r_{\rm disk}$, the mass of the central protostar \red{$M_{*}$} and the internal dust density $\rho_{\rm mat}$. Our results are well summarized in equation (\ref{33}).

\red{Our results show that the dust grains with size of $100 \rm \mu m $ to $ 1 \rm mm$ can be \red{lifted} by the magnetically driven outflow and can distribute in the well outside of the disk, (i.e., the envelope region) as shown in equation (\ref{33}) if the mass ejection rate of the outflow have the value of $ 10^{-7}\,M_{\odot}\,\rm yr^{-1}$ to $10^{-6}\,M_{\odot}\,\rm yr^{-1}$. It also shows that the maximum size of dust strongly depends on the mass ejection rate.}
 
There is an interesting consistency between our results and those indicated by recent (sub)millimeter multi-wavelength observations of the dust continuum. \citet{2024ApJ...961...90C} show a correlation between the mass ejection rate of the outflow and the spectral index $\beta$ of the dust opacity at the envelope scale. Specifically, $\beta$ at the envelope scale is smaller in protostars with stronger outflows. Furthermore, quantitatively, it appears that the outflow mass ejection rate $\dot{M}$ separates objects with and without a significant decrease of $\beta$ at the boundary $\dot{M} \sim 10^{-7},M_{\odot} {\rm yr^{-1}}$ \citep[see their Figure 2 of][]{2024ApJ...961...90C}. Both our results and the recent observations suggest a possible scenario that explains the presence of large dust grains in the envelope, namely that the dust grains grown in the disk are lifted by the outflow and transported to the envelope, resulting in the presence of the grown dust in the envelope. We named this phenomenon as ``Ashfall" \citep[see,][]{2021ApJ...920L..35T} and \citep{2024ApJ...961...90C} named this phenomenon as ``Chimney Flues".

There are several caveats and open questions to be resolved in the future. First, it needs to be clarified whether the outflow can supply sufficient amount of mm-sized the dust grains to decrease the $\beta$ in the envelopes. This needs to be investigated with long-term multi-dimensional MHD simulations to determine how long and how strong the outflow will continue.

\red{
In this study, we assume that dust moves along magnetic field lines.
However, when the poloidal drift velocity is large, the drift velocity perpendicular to the magnetic field lines $v_{\rm drift}$ may also become significant. In that case, dust will cross magnetic field lines as it rises.
 The degree of displacement in the drift direction $\Delta_{\rm drift}$ relative to the displacement in the poloidal direction can be estimated as (\ref{drif}),}
\red{
\begin{equation}
\label{drif}
    \frac{\Delta_{\rm drift}}{\sqrt{r^{2}+z^{2}}} \sim \frac{\Delta_{\rm drift}}{z} \sim \frac{\Delta_{\rm drift}/t_{\rm dyn2}}{z/t_{\rm dyn2}} \sim \frac{v_{\rm drift}}{v_{d,p}},
\end{equation}}
\red{where $v_{\rm drift}$ is given as,}

\begin{align}
    \textcolor{black}{v_{\rm drift} =} \textcolor{black}{t_{\rm stop}\bigg(\frac{GM}{(r^{2}+z^{2})^{3/2}}(z\cos\alpha-r\sin\alpha ) 
    + \frac{v_{\phi}^{2}}{r}\sin\alpha\bigg)}. \label{dust-pol3}
\end{align}

\red{ Figure \ref{vd} shows the poloidal dust velocity $v_{d,p}$ (dashed line) and the drift dust velocity $v_{\rm drift}$ (solid line) with the different dust sizes. From the figure \ref{vd}, it shows that $v_{\rm drift}$ is less than $v_{ d,p}$ up to the escape point for the dust grains with the size of 1 mm or less. This suggests that the degree of displacement in the drift direction relative to the displacement in the poloidal direction $\Delta_{\rm drift}/\sqrt{r^{2}+z^{2}}$ may not be significant even for the 1 mm sized dust grains in our fiducial model. 
Nevertheless, the displacement perpendicular to the field lines can potentially be important, which cannot be captured by our one-dimensional model. Therefore, further investigation using multi-dimensional simulations is also important in future studies.}

\red{Without disk turbulence, it is expected that the (sub)millimeter-sized dust grains tend to settle toward the disk midplane. However, our study suggests that the upward (laminal) flow from the midplane to the disk surface can lift the (sub)millimeter-sized dust grains to upper layers (see section \ref{lift}). As a result, the dust grains can avoid settling to the disk midplane even without the turbulence.}

Although the present study analyzed the dust dynamics with a fixed dust size, the dust grains with different dust sizes can have large relative velocities in the outflow.  It will be important for future studies to consider the possibility and consequences of coalescences and disruptions due to the dust collision in the outflow.

\section*{Acknowledgment}
We thank the anonymous referee for the insightful comments.
This study is supported by JST FOREST Program, Grant Number JPMJFR2234.

\bibliographystyle{aasjournal}
\bibliography{export-bibtex(9)}{}

\begin{thebibliography}{}
\expandafter\ifx\csname natexlab\endcsname\relax\def\natexlab#1{#1}\fi
\providecommand{\url}[1]{\href{#1}{#1}}
\providecommand{\dodoi}[1]{doi:~\href{http://doi.org/#1}{\nolinkurl{#1}}}
\providecommand{\doeprint}[1]{\href{http://ascl.net/#1}{\nolinkurl{http://ascl.net/#1}}}
\providecommand{\doarXiv}[1]{\href{https://arxiv.org/abs/#1}{\nolinkurl{https://arxiv.org/abs/#1}}}

\bibitem[{{Bai} {et~al.}(2016){Bai}, {Ye}, {Goodman}, \&
  {Yuan}}]{2016ApJ...818..152B}
{Bai}, X.-N., {Ye}, J., {Goodman}, J., \& {Yuan}, F. 2016, \apj, 818, 152,
  \dodoi{10.3847/0004-637X/818/2/152}

\bibitem[{{Blandford} \& {Payne}(1982)}]{1982MNRAS.199..883B}
{Blandford}, R.~D., \& {Payne}, D.~G. 1982, \mnras, 199, 883

\bibitem[{{Cacciapuoti} {et~al.}(2024){Cacciapuoti}, {Testi}, {Podio},
  {Codella}, {Maury}, {De Simone}, {Hennebelle}, {Lebreuilly}, {Klessen}, \&
  {Molinari}}]{2024ApJ...961...90C}
{Cacciapuoti}, L., {Testi}, L., {Podio}, L., {et~al.} 2024, \apj, 961, 90,
  \dodoi{10.3847/1538-4357/ad0f17}

\bibitem[{{Cao} \& {Spruit}(1994)}]{1994A&A...287...80C}
{Cao}, X., \& {Spruit}, H.~C. 1994, \aap, 287, 80

\bibitem[{{Carrasco-Gonz{\'a}lez} {et~al.}(2019){Carrasco-Gonz{\'a}lez},
  {Sierra}, {Flock}, {Zhu}, {Henning}, {Chandler}, {Galv{\'a}n-Madrid},
  {Mac{\'\i}as}, {Anglada}, {Linz}, {Osorio}, {Rodr{\'\i}guez}, {Testi},
  {Torrelles}, {P{\'e}rez}, \& {Liu}}]{2019ApJ...883...71C}
{Carrasco-Gonz{\'a}lez}, C., {Sierra}, A., {Flock}, M., {et~al.} 2019, \apj,
  883, 71, \dodoi{10.3847/1538-4357/ab3d33}

\bibitem[{{Chiang} {et~al.}(2012){Chiang}, {Looney}, \&
  {Tobin}}]{2012ApJ...756..168C}
{Chiang}, H.-F., {Looney}, L.~W., \& {Tobin}, J.~J. 2012, \apj, 756, 168,
  \dodoi{10.1088/0004-637X/756/2/168}

\bibitem[{{D'Alessio} {et~al.}(2001){D'Alessio}, {Calvet}, \&
  {Hartmann}}]{2001ApJ...553..321D}
{D'Alessio}, P., {Calvet}, N., \& {Hartmann}, L. 2001, \apj, 553, 321,
  \dodoi{10.1086/320655}

\bibitem[{{Draine}(2006)}]{2006ApJ...636.1114D}
{Draine}, B.~T. 2006, \apj, 636, 1114, \dodoi{10.1086/498130}

\bibitem[{{Epstein}(1924)}]{1924PhRv...23..710E}
{Epstein}, P.~S. 1924, Physical Review, 23, 710, \dodoi{10.1103/PhysRev.23.710}

\bibitem[{{Finkbeiner} {et~al.}(1999){Finkbeiner}, {Davis}, \&
  {Schlegel}}]{1999ApJ...524..867F}
{Finkbeiner}, D.~P., {Davis}, M., \& {Schlegel}, D.~J. 1999, \apj, 524, 867,
  \dodoi{10.1086/307852}

\bibitem[{{Galametz} {et~al.}(2019){Galametz}, {Maury}, {Valdivia}, {Testi},
  {Belloche}, \& {Andr{\'e}}}]{2019A&A...632A...5G}
{Galametz}, M., {Maury}, A.~J., {Valdivia}, V., {et~al.} 2019, \aap, 632, A5,
  \dodoi{10.1051/0004-6361/201936342}

\bibitem[{{Guillet} {et~al.}(2020){Guillet}, {Hennebelle}, {Pineau des
  For{\^e}ts}, {Marcowith}, {Commer{\c{c}}on}, \&
  {Marchand}}]{2020A&A...643A..17G}
{Guillet}, V., {Hennebelle}, P., {Pineau des For{\^e}ts}, G., {et~al.} 2020,
  \aap, 643, A17, \dodoi{10.1051/0004-6361/201937387}

\bibitem[{{J{\o}rgensen} {et~al.}(2007){J{\o}rgensen}, {Bourke}, {Myers}, {Di
  Francesco}, {van Dishoeck}, {Lee}, {Ohashi}, {Sch{\"o}ier}, {Takakuwa},
  {Wilner}, \& {Zhang}}]{2007ApJ...659..479J}
{J{\o}rgensen}, J.~K., {Bourke}, T.~L., {Myers}, P.~C., {et~al.} 2007, \apj,
  659, 479, \dodoi{10.1086/512230}

\bibitem[{{Kudoh} \& {Shibata}(1997)}]{1997ApJ...474..362K}
{Kudoh}, T., \& {Shibata}, K. 1997, \apj, 474, 362, \dodoi{10.1086/303437}

\bibitem[{{Kwon} {et~al.}(2009){Kwon}, {Looney}, {Mundy}, {Chiang}, \&
  {Kemball}}]{2009ApJ...696..841K}
{Kwon}, W., {Looney}, L.~W., {Mundy}, L.~G., {Chiang}, H.-F., \& {Kemball},
  A.~J. 2009, \apj, 696, 841, \dodoi{10.1088/0004-637X/696/1/841}

\bibitem[{{Kwon} {et~al.}(2015){Kwon}, {Looney}, {Mundy}, \&
  {Welch}}]{2015ApJ...808..102K}
{Kwon}, W., {Looney}, L.~W., {Mundy}, L.~G., \& {Welch}, W.~J. 2015, \apj, 808,
  102, \dodoi{10.1088/0004-637X/808/1/102}

\bibitem[{{Laibe} \& {Price}(2012)}]{2012MNRAS.420.2365L}
{Laibe}, G., \& {Price}, D.~J. 2012, \mnras, 420, 2365,
  \dodoi{10.1111/j.1365-2966.2011.20201.x}

\bibitem[{{Lebreuilly} {et~al.}(2020){Lebreuilly}, {Commer{\c{c}}on}, \&
  {Laibe}}]{2020A&A...641A.112L}
{Lebreuilly}, U., {Commer{\c{c}}on}, B., \& {Laibe}, G. 2020, \aap, 641, A112,
  \dodoi{10.1051/0004-6361/202038174}

\bibitem[{{Lebreuilly} {et~al.}(2023){Lebreuilly}, {Vallucci-Goy}, {Guillet},
  {Lombart}, \& {Marchand}}]{2023MNRAS.518.3326L}
{Lebreuilly}, U., {Vallucci-Goy}, V., {Guillet}, V., {Lombart}, M., \&
  {Marchand}, P. 2023, \mnras, 518, 3326, \dodoi{10.1093/mnras/stac3220}

\bibitem[{{Miotello} {et~al.}(2014){Miotello}, {Testi}, {Lodato}, {Ricci},
  {Rosotti}, {Brooks}, {Maury}, \& {Natta}}]{2014A&A...567A..32M}
{Miotello}, A., {Testi}, L., {Lodato}, G., {et~al.} 2014, \aap, 567, A32,
  \dodoi{10.1051/0004-6361/201322945}

\bibitem[{{Ormel} {et~al.}(2009){Ormel}, {Paszun}, {Dominik}, \&
  {Tielens}}]{2009A&A...502..845O}
{Ormel}, C.~W., {Paszun}, D., {Dominik}, C., \& {Tielens}, A.~G.~G.~M. 2009,
  \aap, 502, 845, \dodoi{10.1051/0004-6361/200811158}

\bibitem[{{P{\'e}rez} {et~al.}(2015){P{\'e}rez}, {Chandler}, {Isella},
  {Carpenter}, {Andrews}, {Calvet}, {Corder}, {Deller}, {Dullemond}, {Greaves},
  {Harris}, {Henning}, {Kwon}, {Lazio}, {Linz}, {Mundy}, {Ricci}, {Sargent},
  {Storm}, {Tazzari}, {Testi}, \& {Wilner}}]{2015ApJ...813...41P}
{P{\'e}rez}, L.~M., {Chandler}, C.~J., {Isella}, A., {et~al.} 2015, \apj, 813,
  41, \dodoi{10.1088/0004-637X/813/1/41}

\bibitem[{{Planck Collaboration} {et~al.}(2014{\natexlab{a}}){Planck
  Collaboration}, {Abergel}, {Ade}, {Aghanim}, {Alves}, {Aniano}, {Arnaud},
  {Ashdown}, {Aumont}, {Baccigalupi}, {Banday}, {Barreiro}, {Bartlett},
  {Battaner}, {Benabed}, {Benoit-L{\'e}vy}, {Bernard}, {Bersanelli},
  {Bielewicz}, {Bobin}, {Bonaldi}, {Bond}, {Bouchet}, {Boulanger}, {Burigana},
  {Cardoso}, {Catalano}, {Chamballu}, {Chiang}, {Christensen}, {Clements},
  {Colombi}, {Colombo}, {Couchot}, {Crill}, {Cuttaia}, {Danese}, {Davis}, {de
  Bernardis}, {de Rosa}, {de Zotti}, {Delabrouille}, {D{\'e}sert}, {Dickinson},
  {Diego}, {Dole}, {Donzelli}, {Dor{\'e}}, {Douspis}, {Dupac}, {Efstathiou},
  {En{\ss}lin}, {Eriksen}, {Falgarone}, {Finelli}, {Forni}, {Frailis},
  {Franceschi}, {Galeotta}, {Ganga}, {Ghosh}, {Giard}, {Giraud-H{\'e}raud},
  {Gonz{\'a}lez-Nuevo}, {G{\'o}rski}, {Gregorio}, {Gruppuso}, {Guillet},
  {Hansen}, {Harrison}, {Helou}, {Henrot-Versill{\'e}},
  {Hern{\'a}ndez-Monteagudo}, {Herranz}, {Hildebrandt}, {Hivon}, {Hobson},
  {Holmes}, {Hornstrup}, {Hovest}, {Huffenberger}, {Jaffe}, {Jaffe}, {Joncas},
  {Jones}, {Jones}, {Juvela}, {Kalberla}, {Keih{\"a}nen}, {Kerp}, {Keskitalo},
  {Kisner}, {Kneissl}, {Knoche}, {Kunz}, {Kurki-Suonio}, {Lagache},
  {L{\"a}hteenm{\"a}ki}, {Lamarre}, {Lasenby}, {Lawrence}, {Leonardi},
  {Levrier}, {Liguori}, {Lilje}, {Linden-V{\o}rnle}, {L{\'o}pez-Caniego},
  {Lubin}, {Mac{\'\i}as-P{\'e}rez}, {Maffei}, {Maino}, {Mandolesi}, {Maris},
  {Marshall}, {Martin}, {Mart{\'\i}nez-Gonz{\'a}lez}, {Masi}, {Massardi},
  {Matarrese}, {Mazzotta}, {Melchiorri}, {Mendes}, {Mennella}, {Migliaccio},
  {Mitra}, {Miville-Desch{\^e}nes}, {Moneti}, {Montier}, {Morgante},
  {Mortlock}, {Munshi}, {Murphy}, {Naselsky}, {Nati}, {Natoli}, {Noviello},
  {Novikov}, {Novikov}, {Oxborrow}, {Pagano}, {Pajot}, {Paoletti}, {Pasian},
  {Perdereau}, {Perotto}, {Perrotta}, {Piacentini}, {Piat}, {Pierpaoli},
  {Pietrobon}, {Plaszczynski}, {Pointecouteau}, {Polenta}, {Ponthieu}, {Popa},
  {Pratt}, {Prunet}, {Puget}, {Rachen}, {Reach}, {Rebolo}, {Reinecke},
  {Remazeilles}, {Renault}, {Ricciardi}, {Riller}, {Ristorcelli}, {Rocha},
  {Rosset}, {Roudier}, {Rusholme}, {Sandri}, {Savini}, {Spencer}, {Starck},
  {Sureau}, {Sutton}, {Suur-Uski}, {Sygnet}, {Tauber}, {Terenzi}, {Toffolatti},
  {Tomasi}, {Tristram}, {Tucci}, {Umana}, {Valenziano}, {Valiviita}, {Van
  Tent}, {Verstraete}, {Vielva}, {Villa}, {Wade}, {Wandelt}, {Winkel}, {Yvon},
  {Zacchei}, \& {Zonca}}]{2014A&A...566A..55P}
{Planck Collaboration}, {Abergel}, A., {Ade}, P.~A.~R., {et~al.}
  2014{\natexlab{a}}, \aap, 566, A55, \dodoi{10.1051/0004-6361/201323270}

\bibitem[{{Planck Collaboration} {et~al.}(2014{\natexlab{b}}){Planck
  Collaboration}, {Abergel}, {Ade}, {Aghanim}, {Alves}, {Aniano},
  {Armitage-Caplan}, {Arnaud}, {Ashdown}, {Atrio-Barandela}, {Aumont},
  {Baccigalupi}, {Banday}, {Barreiro}, {Bartlett}, {Battaner}, {Benabed},
  {Beno{\^\i}t}, {Benoit-L{\'e}vy}, {Bernard}, {Bersanelli}, {Bielewicz},
  {Bobin}, {Bock}, {Bonaldi}, {Bond}, {Borrill}, {Bouchet}, {Boulanger},
  {Bridges}, {Bucher}, {Burigana}, {Butler}, {Cardoso}, {Catalano},
  {Chamballu}, {Chary}, {Chiang}, {Chiang}, {Christensen}, {Church}, {Clemens},
  {Clements}, {Colombi}, {Colombo}, {Combet}, {Couchot}, {Coulais}, {Crill},
  {Curto}, {Cuttaia}, {Danese}, {Davies}, {Davis}, {de Bernardis}, {de Rosa},
  {de Zotti}, {Delabrouille}, {Delouis}, {D{\'e}sert}, {Dickinson}, {Diego},
  {Dole}, {Donzelli}, {Dor{\'e}}, {Douspis}, {Draine}, {Dupac}, {Efstathiou},
  {En{\ss}lin}, {Eriksen}, {Falgarone}, {Finelli}, {Forni}, {Frailis},
  {Fraisse}, {Franceschi}, {Galeotta}, {Ganga}, {Ghosh}, {Giard}, {Giardino},
  {Giraud-H{\'e}raud}, {Gonz{\'a}lez-Nuevo}, {G{\'o}rski}, {Gratton},
  {Gregorio}, {Grenier}, {Gruppuso}, {Guillet}, {Hansen}, {Hanson}, {Harrison},
  {Helou}, {Henrot-Versill{\'e}}, {Hern{\'a}ndez-Monteagudo}, {Herranz},
  {Hildebrandt}, {Hivon}, {Hobson}, {Holmes}, {Hornstrup}, {Hovest},
  {Huffenberger}, {Jaffe}, {Jaffe}, {Jewell}, {Joncas}, {Jones}, {Juvela},
  {Keih{\"a}nen}, {Keskitalo}, {Kisner}, {Knoche}, {Knox}, {Kunz},
  {Kurki-Suonio}, {Lagache}, {L{\"a}hteenm{\"a}ki}, {Lamarre}, {Lasenby},
  {Laureijs}, {Lawrence}, {Leonardi}, {Le{\'o}n-Tavares}, {Lesgourgues},
  {Levrier}, {Liguori}, {Lilje}, {Linden-V{\o}rnle}, {L{\'o}pez-Caniego},
  {Lubin}, {Mac{\'\i}as-P{\'e}rez}, {Maffei}, {Maino}, {Mandolesi}, {Maris},
  {Marshall}, {Martin}, {Mart{\'\i}nez-Gonz{\'a}lez}, {Masi}, {Massardi},
  {Matarrese}, {Matthai}, {Mazzotta}, {McGehee}, {Melchiorri}, {Mendes},
  {Mennella}, {Migliaccio}, {Mitra}, {Miville-Desch{\^e}nes}, {Moneti},
  {Montier}, {Morgante}, {Mortlock}, {Munshi}, {Murphy}, {Naselsky}, {Nati},
  {Natoli}, {Netterfield}, {N{\o}rgaard-Nielsen}, {Noviello}, {Novikov},
  {Novikov}, {Osborne}, {Oxborrow}, {Paci}, {Pagano}, {Pajot}, {Paladini},
  {Paoletti}, {Pasian}, {Patanchon}, {Perdereau}, {Perotto}, {Perrotta},
  {Piacentini}, {Piat}, {Pierpaoli}, {Pietrobon}, {Plaszczynski},
  {Pointecouteau}, {Polenta}, {Ponthieu}, {Popa}, {Poutanen}, {Pratt},
  {Pr{\'e}zeau}, {Prunet}, {Puget}, {Rachen}, {Reach}, {Rebolo}, {Reinecke},
  {Remazeilles}, {Renault}, {Ricciardi}, {Riller}, {Ristorcelli}, {Rocha},
  {Rosset}, {Roudier}, {Rowan-Robinson}, {Rubi{\~n}o-Mart{\'\i}n}, {Rusholme},
  {Sandri}, {Santos}, {Savini}, {Scott}, {Seiffert}, {Shellard}, {Spencer},
  {Starck}, {Stolyarov}, {Stompor}, {Sudiwala}, {Sunyaev}, {Sureau}, {Sutton},
  {Suur-Uski}, {Sygnet}, {Tauber}, {Tavagnacco}, {Terenzi}, {Toffolatti},
  {Tomasi}, {Tristram}, {Tucci}, {Tuovinen}, {T{\"u}rler}, {Umana},
  {Valenziano}, {Valiviita}, {Van Tent}, {Verstraete}, {Vielva}, {Villa},
  {Vittorio}, {Wade}, {Wandelt}, {Welikala}, {Ysard}, {Yvon}, {Zacchei}, \&
  {Zonca}}]{2014A&A...571A..11P}
---. 2014{\natexlab{b}}, \aap, 571, A11, \dodoi{10.1051/0004-6361/201323195}

\bibitem[{{Ricci} {et~al.}(2010){Ricci}, {Testi}, {Natta}, {Neri}, {Cabrit}, \&
  {Herczeg}}]{2010A&A...512A..15R}
{Ricci}, L., {Testi}, L., {Natta}, A., {et~al.} 2010, \aap, 512, A15,
  \dodoi{10.1051/0004-6361/200913403}

\bibitem[{{Sakurai}(1985)}]{1985A&A...152..121S}
{Sakurai}, T. 1985, \aap, 152, 121

\bibitem[{{Sakurai}(1987)}]{1987PASJ...39..821S}
---. 1987, \pasj, 39, 821

\bibitem[{{Silsbee} {et~al.}(2022){Silsbee}, {Akimkin}, {Ivlev}, {Testi},
  {Gong}, \& {Caselli}}]{2022ApJ...940..188S}
{Silsbee}, K., {Akimkin}, V., {Ivlev}, A.~V., {et~al.} 2022, \apj, 940, 188,
  \dodoi{10.3847/1538-4357/ac978b}

\bibitem[{{Tazzari} {et~al.}(2016){Tazzari}, {Testi}, {Ercolano}, {Natta},
  {Isella}, {Chandler}, {P{\'e}rez}, {Andrews}, {Wilner}, {Ricci}, {Henning},
  {Linz}, {Kwon}, {Corder}, {Dullemond}, {Carpenter}, {Sargent}, {Mundy},
  {Storm}, {Calvet}, {Greaves}, {Lazio}, \& {Deller}}]{2016A&A...588A..53T}
{Tazzari}, M., {Testi}, L., {Ercolano}, B., {et~al.} 2016, \aap, 588, A53,
  \dodoi{10.1051/0004-6361/201527423}

\bibitem[{{Toomre}(1964)}]{1964ApJ...139.1217T}
{Toomre}, A. 1964, \apj, 139, 1217, \dodoi{10.1086/147861}

\bibitem[{{Tsukamoto} {et~al.}(2021{\natexlab{a}}){Tsukamoto}, {Machida}, \&
  {Inutsuka}}]{2021ApJ...913..148T}
{Tsukamoto}, Y., {Machida}, M.~N., \& {Inutsuka}, S. 2021{\natexlab{a}}, \apj,
  913, 148, \dodoi{10.3847/1538-4357/abf5db}

\bibitem[{{Tsukamoto} {et~al.}(2021{\natexlab{b}}){Tsukamoto}, {Machida}, \&
  {Inutsuka}}]{2021ApJ...920L..35T}
{Tsukamoto}, Y., {Machida}, M.~N., \& {Inutsuka}, S.-i. 2021{\natexlab{b}},
  \apjl, 920, L35, \dodoi{10.3847/2041-8213/ac2b2f}

\bibitem[{{Valdivia} {et~al.}(2019){Valdivia}, {Maury}, {Brauer}, {Hennebelle},
  {Galametz}, {Guillet}, \& {Reissl}}]{2019MNRAS.488.4897V}
{Valdivia}, V., {Maury}, A., {Brauer}, R., {et~al.} 2019, \mnras, 488, 4897,
  \dodoi{10.1093/mnras/stz2056}

\bibitem[{{Vorobyov} \& {Elbakyan}(2019)}]{2019A&A...631A...1V}
{Vorobyov}, E.~I., \& {Elbakyan}, V.~G. 2019, \aap, 631, A1,
  \dodoi{10.1051/0004-6361/201936132}

\bibitem[{{Vorobyov} {et~al.}(2019){Vorobyov}, {Skliarevskii}, {Elbakyan},
  {Pavlyuchenkov}, {Akimkin}, \& {Guedel}}]{2019A&A...627A.154V}
{Vorobyov}, E.~I., {Skliarevskii}, A.~M., {Elbakyan}, V.~G., {et~al.} 2019,
  \aap, 627, A154, \dodoi{10.1051/0004-6361/201935438}

\bibitem[{{Weingartner} \& {Draine}(2001)}]{2001ApJ...548..296W}
{Weingartner}, J.~C., \& {Draine}, B.~T. 2001, \apj, 548, 296,
  \dodoi{10.1086/318651}

\bibitem[{{Wong} {et~al.}(2016){Wong}, {Hirashita}, \&
  {Li}}]{2016PASJ...68...67W}
{Wong}, Y. H.~V., {Hirashita}, H., \& {Li}, Z.-Y. 2016, \pasj, 68, 67,
  \dodoi{10.1093/pasj/psw066}

\end{thebibliography}
\end{document}